\documentclass[11pt,draftcls,onecolumn]{IEEEtran}
\usepackage{amssymb, amsmath, graphicx, cite, color, epsfig}

\newcommand{\integers}{\mathbb{Z}}

\newcommand{\complex}{\mathbb{C}}

\newcommand{\Fig}   {\mbox{Fig.} }

\newcommand{\eqdef}{ := }

\newcommand{\thmend}{\hspace*{\fill}~\QEDopen\par\endtrivlist\unskip}

\newcommand{\na} { {\sf a} }
\newcommand{\nb} { {\sf b} }
\newcommand{\nr} { {\sf r} }

\newcommand{\prot}{ {\sf P} }
\newcommand{\pa} { {(1)} }
\newcommand{\pb} { {(2)} }
\newcommand{\pc} { {(3)} }

\newtheorem{theorem}{Theorem}
\newtheorem{lemma}[theorem]{Lemma}

\newtheorem{definition}[theorem]{Definition}
\newtheorem{corollary}[theorem]{Corollary}

\allowdisplaybreaks[1]

\begin{document}
\title{Bi-directional half-duplex protocols with multiple relays}
\author{Sang Joon Kim, Natasha Devroye,
and Vahid Tarokh
\thanks{ Sang Joon Kim and Natasha Devroye were, and Vahid Tarokh is with the School
of Engineering and Applied Sciences, Harvard University, Cambridge,
MA 02138. Natasha Devroye is currently with the University of Illinois at Chicago, Chicago, IL 60607.
Emails:~adella0919@gmail.com, devroye@ece.uic.edu,
vahid@seas.harvard.edu.
This research is supported in part by NSF grant number ACI-0330244 and ARO MURI grant number W911NF-07-1-0376. This work was supported in part by the Army Research Office, under the MURI award No. N00014-01-1-0859.  The views expressed in this paper are
those of the authors alone and not of the sponsor.
} }

\maketitle

\begin{abstract}
In a bi-directional relay channel, two nodes wish to exchange independent messages over a shared wireless half-duplex channel with the help of relays. Recent work has considered information theoretic limits of the bi-directional relay channel with a single relay. In this work we consider bi-directional relaying with multiple relays. We derive achievable rate regions and outer bounds for half-duplex protocols with multiple decode and forward relays and compare these to the same protocols with amplify and forward relays in an additive white Gaussian noise channel. We consider three novel classes of half-duplex protocols: the $(m,2)$  2 phase protocol with $m$ relays, the $(m,3)$ 3 phase protocol with $m$ relays, and general $(m, t)$ Multiple Hops and Multiple Relays (MHMR) protocols, where $m$ is the total number of relays and $3<t\leq m+2$ is the number of temporal phases in the protocol. The $(m,2)$ and $(m,3)$ protocols extend previous bi-directional relaying protocols for a single $m=1$ relay, while the new  $(m,t)$ protocol efficiently combines multi-hop routing with message-level network coding. Finally, we provide a comprehensive treatment of the MHMR protocols with decode and forward relaying and amplify and forward relaying in the Gaussian noise, obtaining their respective achievable rate regions, outer bounds and relative performance under different SNRs and relay geometries, including an analytical comparison on the protocols at low and high SNR.
\end{abstract}



\section{Introduction}

\label{sec:intro}

In bi-directional channels, two \emph{terminal nodes}  ($\na$ and $\nb$) wish to exchange independent messages.  The ``two-way channel'' was first considered in \cite{Shannon:1961}, where full-duplex operation in which nodes may transmit and receive simultaneously was assumed. While the capacity region is known for ``restricted'' and additive white Gaussian noise two-way channels, it remains unknown in general.
In wireless channels or mesh networks, two-way or bi-directional communication may take place with the help of $m$ other nodes $\nr_i$, $i\in \{ 1,2, \cdots m\}$  termed relays.
Since full-duplex operation is, with current technology, of limited practical significance, in this work we assume that the nodes are {\em half-duplex}, i.e. at each point in time, a node can either transmit or receive symbols, but not both. Consequently, nodes communicate according to pre-defined ``protocols'' which indicate which node transmits when.   

Our main goal is to determine the limits of bi-directional communication with multiple relays. To do so, we propose and determine the achievable rate regions, as well as outer bounds obtained using several protocols.
The protocols we propose for the multiple-relay bi-directional channel may be described in terms of two parameters: the number of relays, $m$, and the number of temporal phases $t$, called \emph{hops}. Throughout this work, \emph{phases} and \emph{hops} are used interchangeably.  We also define an \emph{intermediate hop} as a hop in which only relays transmit (and not the terminal nodes). Note that our protocols are all composed of a number of temporal phases/hops  due to the half-duplex nature of the channel. 

{\bf Protocols.} We denote our proposed protocols as $(m,t)$ MHMR (Multiple Hops and Multiple Relays) protocols, for general positive integers $m\geq 2$ and $t\geq 2$.
For the special case of two hops ($t=2$),  the terminal nodes may  simultaneously transmit in phase 1 as in the MABC (Multiple Access Broadcast Channel) protocol of \cite{SKim:2007}, while the relays transmit the decoded messages to the terminal nodes in phase 2. For the special case of three hops ($t=3$) the terminal nodes may sequentially transmit in the first two phases as in the TDBC (Time Division Broadcast Channel) protocol of \cite{SKim:2007}, after which the relays transmit in phase 3. However, the $(m,t)$ MHMR (Multiple Hops and Multiple Relays) protocol for $t>3$ is not an immediate extension/generalization of the $(m,2)$ and $(m,3)$ protocols. 

{\bf Relaying scheme.} While a protocol in this work defines the temporal aspect (phases) of bi-directional communication, it does not specify the type of relaying a node may perform, or \emph{relaying scheme}. That is,
for each of the MHMR protocols, the relays may process and
forward the received signals differently. Standard forwarding techniques include decode-and-forward, amplify-and-forward, compress-and-forward, and de-noise and forward. We consider only the first two relaying schemes. In the {\em Decode and Forward} (DF) scheme, the relays decode  messages from the other nodes before re-encoding them for transmission. The DF scheme requires  the full codebooks of all nodes and a large amount of computation at the relays $\{\nr_i\}$.
In the {\em Amplify and Forward} (AF) scheme, the relays $\{\nr_i\}$ construct their symbol by symbol replications of the received symbols. The AF scheme requires very little computation, and carries the noise incurred in the first stage(s) forward during the latter relaying stage(s). The relative benefits and merits of the temporal protocols
and relaying schemes are summarized in Tables
\ref{table:protocol} and  \ref{table:relaying} where we compare the amount of knowledge/computation at the relays as well as the amount of interference and side information present. By side information we mean information obtained from the wireless channel in a particular phase which may be combined with information obtained in different stages to potentially improve decoding or increase transmission rates.

\subsection{Prior related work}
The work extends upon the large body of work on the {\it bi-directional relay} or {\it two-way relay channel,} which has been considered from a variety of perspectives for a channel with a {\it single} relay.  While we cannot enumerate all work in the area, it may be differentiated roughly based on combinations of assumptions that are made on the type of relaying (CF, DF, AF, de-noise, mixed, lattice codes), on the duplex abilities of nodes (half-duplex or full-duplex), and on the presence/absence of direct links between terminal nodes.  We highlight examples of work in this area before  describing extensions to multiple relay nodes.

1. {\bf Relaying type:} Amplify-and-forward (AF) is the simplest of relaying types, in which relays simply re-scale and re-transmit the received signal. Its benefits lie in its simplicity and lack of relay-node processing, and as such AF-relaying schemes are
 used as references against which to compare the performance of more complex relaying schemes \cite{SKim:2007, Rankov:2007,Popovski:ICC, Popovski:2006a, Popovski:2006b, Ho:2008}. In Decode and forward (DF) relaying  the relay decodes all messages before re-transmitting them.
  Examples of work which assume bi-directional DF relaying include \cite{Larsson:2005, Larsson:2006, Oechtering:2007, Oechtering:2008, Oechtering:2009, OB08pacm, Wyrembelski:2008, SKim:2007, Kim:ISIT2009, Schnurr:2007, Kim:sarnoff, Rankov:2006}.  To include network coding at the message level (rather than at the signal level in analog network coding \cite{Katti:2007}) decode and forward relays must be assumed. DF relaying allows for network coding at the message level for the broadcast phase and prevents the re-transmission and possible amplification of noise. These come at the cost of {\it requiring} the relays to decode the messages, possibly reducing the permissible transmission rates. Compress and forward (CF), as first introduced in the context of the classical relay channel \cite{Cover:2006}, and considered in the bi-directional relaying context in \cite{Rankov:2006, Schnurr:2007, Kim:sarnoff, gunduz2008rate} and the conceptually related de-noise and forward  \cite{Popovski:ICC}, \cite{Popovski:2006a}, \cite{Popovski:2006b}, requires the relay to re-transmit a quantized or compressed version of the received signal. This scheme has the advantage that the rate need not be lowered so as to allow the relay to fully decode it, but may still mitigate some of the noise amplification effects seen in AF relaying by judicious choice of the quantizer or compressor.
   An elegant and powerful recent extension of compress-and-forward type schemes  is the ``noisy-network coding'' relaying scheme of \cite{lim2010noisy} which involves message repetition coding, relay signal compression, and simultaneous decoding. The general results obtained in \cite{lim2010noisy} include examples of achievable rate regions in special cases including the two-way relay channel.  Finally, in the authors' previous work \cite{Kim:sarnoff, SKim:2007} the {\it mixed} forwarding scheme is also proposed, in which the streams of information traveling in the two directions are treated differently, i.e. one direction may use DF while the other uses CF.  Expanding on the intuition gained from the classical relay channel \cite{Kramer:Gastpar:Gupta}, and as further shown in the bi-directional relaying context \cite{SKim:2007}, when a relay is close (or alternatively sees a weak channel to the source relative to the destination) to the destination it is preferable to perform CF, while if it is closer (or sees a better channel to the source relative to the destination)  to the source DF is a better choice. Thus, mixed schemes in general may be beneficial.

2. {\bf Duplexing:} Both  full-duplex and half-duplex relaying types have been considered in deriving achievable rate regions for for bi-directional relaying. In \cite{Wu:2005, Larsson:2005, Larsson:2006, Kim:ICDCSW07, SKim:2007, Oechtering:2007, Rankov:2007, Popovski:ICC, Popovski:2006a, Popovski:2006b, Vaze:2009, Yuen:VTC, Schnurr:2007, Kim:sarnoff} half-duplex nodes are assumed. This forces communication to take place over a number of phases, using different temporal {\it protocols} which specify which nodes  transmit when  (many may transmit simultaneously creating interference).
Two of the most commonly considered protocols are depicted in \ref{fig:MABC_TDBC}: the  2 phase Multiple-access broadcast channel (MABC) protocol, and   the 3 phase time-division broadcast channel (TDBC) protocol. 
In \cite{Kim:ICDCSW07, SKim:2007} it is shown that neither TDBC or MABC dominate each other for all channel gains and SNRs. In \cite{Kim:sarnoff, SKim:2008a} it is furthermore shown that there is significant interplay between relaying types and protocols: a comprehensive treatment of CF, DF, AF and mixed forwarding schemes are analyzed under  both the MABC and TDBC protocols in \cite{Kim:sarnoff, SKim:2008a}. In \cite{Oechtering:2007, Oechtering:2008, Oechtering:2009, OB08pacm, Wyrembelski:2008} the authors have thoroughly analyzed the broadcast phase from a number of angles and assumptions (thus assuming half-duplex, decode and forward relaying) of the bi-directional relay channel. The full-duplex scenarios have been considered somewhat less: in \cite{Rankov:2006} the authors derived achievable rate regions for the {\it restricted} two-way relay channel using DF, CF and AF schemes, in which the terminals may not cooperate in transmitting their messages. In \cite{nam:2009bit, avestimehr2009capacity} full-duplex nodes are considered in order to analyze the Gaussian noise two-way relay channel from a finite bit perspective.

3. {\bf Direct link between the nodes:} When all communication must pass through the relays, i.e. there are no direct links, the only side-information available at the receiving nodes is that it has a priori: its own message (nothing overheard from previous phases).  Such networks tend to be easier to deal with as the possibility of direct communication between terminal nodes is excluded and thus the tension between how much should be sent through the relay versus how much should be sent directly is eliminated.
While much work has been performed on the bi-directional relay channel in which there is no direct link, there are some notable exceptions including the achievable rate regions in   \cite{Kim:sarnoff, xie2007network, avestimehr2009capacity, Rankov:2006, song:allerton:2010}. We note that while finite-gap results are available for the Gaussian relay channel without direct links \cite{nam:2009bit, avestimehr2009capacity, song:allerton:2010}, similar results hold only for a subset of the channels with direct links \cite{avestimehr2009capacity, song:allerton:2010} (under certain symmetric channel conditions), and capacity remains unknown in general.

{\bf Extensions to multiple relays.} The bi-directional relay channel has been extended to include multiple relays \cite{SKim:2008b, OB08brru, Kim:ISIT2009, Pooniah:2008, Vaze:2008, Vaze:2008codeornot} and multiple terminals nodes (or multiple bi-directional data streams)  \cite{ Ghozlan_MIMO_switch, Chen:2008:CISS,Avestimehr:2009:ITW, Gunduz:2009:ISIT, Kim:ISIT2010}. We focus on prior work in which multiple relays are used to convey a bi-directional information flow.
In \cite{Pooniah:2008} the authors propose an achievable rate region for the two-way
full-duplex two-relay channel using a DF block Markov coding scheme. This paper differs from \cite{Pooniah:2008} in that we consider half-duplex nodes rather than full-duplex nodes and consider an arbitrary number of relays (rather than two).
In \cite{Vaze:2008}
an iterative algorithm is proposed
to achieve the optimal achievable rate region for a MIMO two-way relay channel with multiple relays where each relay employs an amplify and forward (AF)
strategy under both global and local channel state information requirements at all nodes.
The authors furthermore show that  compress and forward strategy is optimal for achieving the  diversity-multiplexing tradeoff for the full-duplex case, in general, and for the half-duplex case in some cases. This again differs from our work in that we assume DF relaying and seek achievable rate regions rather than diversity-multiplexing tradeoffs. In \cite{Vaze:2008codeornot} distributed space-time-coding and end-to-end antenna selection schemes are compared -- mainly in terms of diversity gains --  for one-way and two-way muti-hop relay channels with both full and half-duplex relay nodes. This work differs in that we do not address the relay selection scheme but rather allow all relays to potentially be used, and they may cooperate in a DF fashion,  which will include the rate region achieved using a distributed space-time code. Furthermore, we focus only on achievable rate regions and not on diversity gains.
Finally, in \cite{OB08brru} optimal relay selection  for  unidirectional relay channels \cite{Luo:VTC, Madan:2008, Bletsas:2006} are extended to bi-directional relay networks,  where rate regions which select one relay, or time-share between different two-way relays, are derived. This differs from the work here in that we do not select a single relay for transmission, but rather consider all relays as possible transmitters.

%


\subsection{Contributions and structure}
The main contributions of this work are:
\begin{enumerate}
\item the extension of
previously defined single relay MABC and TDBC protocols to multiple
DF relays,
\item  the introduction of a novel class of general $(m,t)$ MHMR bi-directional DF relaying protocols,
\item the derivation of all associated achievable rate regions and outer
bounds,
\item an analytical comparison of the protocols at asymptotically high and low SNR, using the multiplicative gap as our metric, and
\item  a comprehensive comparison of these schemes with
their AF analogs in Gaussian noise.
\end{enumerate}

Some of the main conclusions
drawn are that, in bi-directional multiple-relay channels, it
may be beneficial to have information flowing in both directions
along a series of hops, where the information is carefully combined
in a network-coding-like fashion. When the number of hops is large
or when the SNR is low, DF outperforms AF as noise is not carried
forward. Furthermore, at high SNR, we analytically show that the $(m,m+2)$ DF MHMR bi-directional protocol scales (in SNR) like the cut-set outer bound, in contrast to the $(m,2)$ and $(m,3)$ protocols. At low SNR, the $(m,m+2)$ DF MHMR similarly outperforms other schemes as its performance does not degrade with the number of relays.
Simulations show that the careful choice of the number of hops and which relays
participate in each hop can lead to significant gains in terms of
the achievable rates.

{\bf Structure.} This paper is structured as follows: in Section \ref{sec:prelim}, we
introduce our notation and review previously determined achievable
rate regions. In Section \ref{sec:protocols}, we introduce novel
$(m,t)$ MHMR protocols. In Section \ref{sec:bounds} we derive
achievable rate regions for the $(m,t)$ MHMR protocols with DF relaying. In Section
\ref{sec:outer_bounds} we derive outer bounds for the MHMR
protocols. In Section \ref{sec:gaussian} we obtain explicit
expressions for achievable rate regions and outer bounds and their  corresponding AF analogs in
Gaussian noise. In Section \ref{sec:analysis}, we analyze the asymptotic performance in the  very low and high SNR regimes. In Section \ref{sec:regions}, we numerically compute
these bounds in the Gaussian noise channel and compare the results
for different powers and channel conditions.

\begin{table}
\caption{Comparison between temporal protocols }
\label{table:protocol}
\centering
\begin{tabular}{c||c|c|c}
  \hline
  Protocol & Side information & Interference & Cooperation\\
  \hline
  $(m,2)$ MABC & not present  & present  & possible\\
  $(m,3)$ TDBC & present  & not present  & possible\\
  $(m,t)$ MHMR ($2< t<m+2$)& accumulated & not present & possible \\
  $(m,m+2)$ MHMR & fully accumulated & not present & impossible\\
  \hline
\end{tabular}
\end{table}

\begin{table}
\caption{Comparison between two relaying schemes }
\label{table:relaying}
\centering
\begin{tabular}{c||c|c|c}
  \hline
  Relaying & Complexity & Noise at relay & Relay needs \\
  \hline
  AF & low & carried  & nothing \\
  DF & high & eliminated & full codebooks \\
  \hline
\end{tabular}
\end{table}


\section{Preliminaries}

\label{sec:prelim}

\subsection{Definitions}
Nodes $\na$ and $\nb$ are the two terminal nodes which wish to exchange their message with the help of the relays in the set ${\cal R}
\eqdef \{\nr_1, \nr_2,\cdots, \nr_m\}$. For
convenience of analysis we define $\nr_0 \eqdef \na$, $\nr_{m+1}
\eqdef \nb$ and use these notations interchangeably in the following
sections.
 The two messages to be transmitted from node $\na$ to node $\nb$ and $\nb$ to $\na$ are independent and denoted by $W_\na
\eqdef W_{\na,\nb}$ and $W_\nb \eqdef W_{\nb,\na}$, of rates $R_\na \eqdef R_{\na,\nb}$ and $R_\nb \eqdef
R_{\nb,\na}$ respectively. That is,  $R_{i,j}$ is used to denote the
transmitted data rate from node $i$ to node $j$, and the message between node $i$ and node $j$, $W_{i,j}$,  and is uniformly distributed in the set $   {\cal
S}_{i,j} \eqdef
\{0, \ldots, \lfloor 2^{nR_{i,j}} \rfloor - 1\}$. Also define ${\cal R}^* \eqdef {\cal R}\cup \{\na,\nb\} =
\{\nr_0,\nr_1,\cdots,\nr_{m+1}\}$ (for $i=\na, j=\nb$ or $i=\nb, j=\na$).
{The non-negative variable $\Delta_\ell$ is defined as the relative time duration of the $\ell^{th}$ phase, where $\sum_\ell \Delta_\ell = 1$. For a given number of total channel uses $n$, $\Delta_{\ell,n}$ is  the duration of the $\ell^{th}$ phase, or equivalently $\Delta_{\ell,n} = \frac{n_{\ell}}{n}$, where $n_\ell$ is the number of channel uses in phase $\ell$. Then $\Delta_{\ell,n} \rightarrow \Delta_{\ell}$ with proper choice of $n_{\ell}$ as $n \rightarrow \infty$.}

For a given block length $n$, we define the transmission random variable at time $1\leq k\leq n$ at
node $i$ by $X_i^k$. Similarly, the received random variable at node $i$ at time $k$ is defined as
$Y_i^k$. The distributions of $X_i^k$ and $Y_i^k$ depend on
the time instance $k$. For all $k$ in phase $\ell$, the distribution of $X_i^k$ are identical and equal to the distribution of the random variable  $X_i^{(\ell)}$ with discrete alphabet ${\cal X}_i^*$ and input distribution $p^{(\ell)}(x_i)$ (and similarly for $Y_i^k$ and $Y_i^{\ell}$). Each node $i$ has channel input alphabet ${\cal X}^*_i = {\cal
X}_i \cup \{ \varnothing \}$ and channel output alphabet ${\cal
Y}^*_i = {\cal Y}_i \cup \{ \varnothing \}$. Because of the
half-duplex constraint, not all nodes transmit/receive during all
phases and we use the dummy symbol $\varnothing$ to denote that
there is no input or no output at a particular node during a
particular phase. {We distinguish the symbol $\varnothing$  from the symbol  $\emptyset$ used to denote the empty set, i.e. $\emptyset = \{\}$.} The half-duplex constraint forces either
$X_i^{(\ell)} = \varnothing$ or $Y_i^{(\ell)} = \varnothing$ in phase $\ell$,  for all
$\ell$ phases.

We then denote $X_S^k \eqdef \{X_i^k | i\in S\}$ as the set of transmissions by all nodes in the set $S$ at time $k$, and $X_{S}^{(\ell)} \eqdef \{X_i^{(\ell)}|i\in S\}$, a set of random variables with channel input distribution $p^{(\ell)}(x_S)$ for phase $\ell$, where $x_S \eqdef \{x_i | i\in S\}$. Lower case letters $x_i$  denotes instances of the upper case $X_i$ which lie in the calligraphic alphabets ${\cal X}_i^*$. Boldface ${\bf x}_i$ is a vector indexed by time at node $i$. For convenience, we drop the notation $\varnothing$ from entropy and the mutual information terms when a node is not transmitting or receiving. For example, $I(X_\na^\pa;Y_\nr^\pa) = I(X_\na^\pa;Y_\nr^\pa|X_\nb^\pa=X_\nr^\pa = \varnothing)$ when $\nb$ and $\nr$ are in receiving mode during phase 1. Finally, we denote ${\bf x}_S \eqdef \{{\bf x}_i | i\in S\}$ as a set of vectors indexed by time. $Q$ denotes a discrete time-sharing random variable with distribution $p(q)$.

The channel is assumed to be discrete and  \emph{memoryless}. For convenience, we drop the notation $\varnothing$
from entropy and mutual information terms when a node is not
transmitting or receiving. Communication takes place over $n$ of channel uses and rates are achieved in the classical
asymptotic sense as $n\rightarrow \infty$.


{ We define $W_{S,T}= \{W_{i,j}| i\in S,~ j\in T ~,~ S,T \subseteq {\cal R}^*\}$.} For a block length $n$, encoders and decoders are functions
${\cal E}: (W_{\{i\},{\cal R}^*}, Y_i^1, \cdots, Y_i^{k-1}) \rightarrow X_i^k$ producing an encoded message, and
${\cal D}: (Y_i^1, \cdots, Y_i^n, W_{\{i\},{\cal R}^*}) \rightarrow \tilde{W}_{j,i}$ producing a decoded message or error, for sending a message from node $j$ to node $i$ at time $k=1,2,\cdots n$.
{ We define the error events $E_{i,j} \eqdef \{W_{i,j} \neq \tilde{W}_{i,j}(.)\}$ for decoding the message $W_{i,j}$
at node $j$ at the end of the block of length $n$, and $E_{i,j}^{(\ell)}$ as the error event at node $j$ in which node $j$ attempts to decode $w_i$ at the end of phase $\ell$ using a joint typicality decoder. Then we define error events $E_{S, T} \eqdef \cup_{i\in S, j\in T} E_{i,j}$ and $E_{S, T}^{(\ell)} \eqdef \cup_{i\in S, j \in T} E_{i,j}^{(\ell)}$. For example, in the $(1,2)$ MABC protocol (single relay two phase protocol), $E_{\na,\nb} = E_{\na,\nr}^\pa \cup E_{\nr,\nb}^\pb$, where
$E_{\na,\nb}$ is the error event when $\na$ sends messages to $\nb$,
$E_{\na,\nr}^\pa$ is the error event when $\na$ sends messages to $\nr$ during phase 1, and
$E_{\nr,\nb}^\pb$ is the error event when $\nr$ sends messages to $\nb$ during phase 2.

Let $A^{(\ell)}(UV)$ represent the set of $\epsilon$-typical
$({\bf u}^{(\ell)},{\bf v}^{(\ell)})$ sequences of length $n \cdot \Delta_{\ell,n}$
according to the distributions $U$ and $V$ in phase $\ell$. The events $D^{(\ell)}({\bf u},{\bf v})\eqdef
\{({\bf u}^{(\ell)},{\bf v}^{(\ell)}) \in A^{(\ell)}(UV)\}$.

A set of rates $R_{i,j}$ is said to be achievable for a protocol with phase durations $\{\Delta_{\ell}\}$ if there exist  encoders/decoders of block length $n = 1, 2, \ldots$   with both $P[E_{i,j}]\rightarrow 0$ and $\Delta_{\ell , n}\rightarrow \Delta_{\ell}$ as $n\rightarrow \infty$ for all $\ell$. An achievable rate region (resp. capacity region) is the closure of a set of (resp. all) achievable rate
tuples for fixed $\{\Delta_{\ell}\}$.

For the $(m,2)$, $(m,3)$ DF MHMR protocols we define $\cal{A}$
(resp. $\cal{B}$) as the set of relays which are able to decode
$w_\na$ (resp. $w_\nb$). We define $I_{S}^{\min}
(X_i^{(\ell)};Y_s^{(\ell)}) \eqdef \min_{s\in S}
I(X_i^{(\ell)};Y_s^{(\ell)})$. For example, $I_{\cal{A}}^{\min}
(X_\na^{\pa};Y_\nr^{\pa}) = \min_{\nr\in \cal{A}}
I(X_\na^{\pa};Y_\nr^{\pa})$, i.e. the minimum mutual information
between node $\na$ and a relay in the set of relays which can decode
$w_\na$. We also define $I_{\o}^{\min}(X;Y)=0$. $\oplus$ denotes an addition in a given additive group $\integers_L$, where $L$ is an arbitrary positive integer and  $\bigotimes$ denotes the Cartesian product, i.e., $\bigotimes_{i=1}^3 {\cal X}_i = {\cal X}_1 \times {\cal X}_2 \times {\cal X}_3 $. }
\subsection{Previous single relay results}

\begin{figure}
\centerline{\epsfig{figure=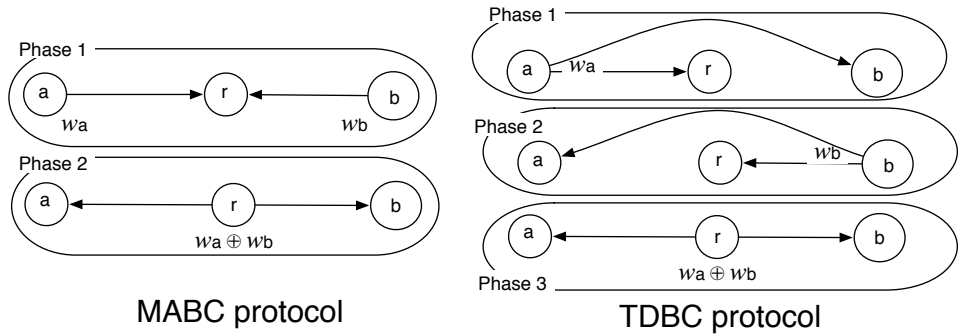, width=13cm}}
\caption{The two phase MABC and three phase TDBC protocols of
\cite{SKim:2007}.} \label{fig:MABC_TDBC}
\end{figure}

We next outline the relevant previously derived \cite{SKim:2007} achievable rate regions of bi-directional decode and forward protocols with a single relay, $\nr$, and terminal nodes $\na$ and $\nb$, as shown in Fig. \ref{fig:MABC_TDBC}. These regions will be used  in the discussions in Sections \ref{sec:protocols} and \ref{sec:bounds}. The three phase protocol is called the {\it Time Division Broadcast} (TDBC) protocol, while the two phase protocol is called the {\it Multiple Access Broadcast} (MABC) protocol.  One of the main conceptual differences between these two protocols is the possibility of \emph{side-information} in the TDBC protocol but not in the MABC protocol.  The two previously considered protocols may be described as:
\begin{enumerate}
\item  TDBC protocol: this consists of the three phases $\na \rightarrow \nr$, $\nb \rightarrow \nr$ and $\na \leftarrow \nr \rightarrow \nb$. In this protocol, only a single node is transmitting at any given point in time. Therefore, by the broadcast nature of the wireless channel, the non-transmitting nodes may listen in and obtain  ``side information'' about the other nodes' transmissions. This may be used to improve the rates of transmission.
\item  MABC protocol: this protocol combines the first two phases of the TDBC protocol and consists of the two phases $\na \rightarrow \nr \leftarrow \nb$ and $\na \leftarrow \nr \rightarrow \nb$. Due to the half-duplex assumption, during phase 1 both source nodes are transmitting and thus cannot obtain any  ``side information'' regarding the other nodes' transmission. It may nonetheless be spectrally efficient since it consists of fewer phases than the TDBC protocol and may take advantage of the multiple-access channel in phase 1.
\end{enumerate}
We now state the results of \cite{SKim:2007} for completeness.

\begin{theorem}
\label{theorem:MABC:one} An achievable rate region of the
half-duplex bi-directional relay channel with the decode and forward
MABC protocol is the closure of the set of all points
$(R_\na,R_\nb)$ satisfying
\begin{align}
R_{\na} &< \min \left\{ \Delta_1 I(X_{\na}^\pa ; Y_{\nr}^\pa | X_{\nb}^\pa,Q),
            \Delta_2 I(X_{\nr}^\pb ; Y_{\nb}^\pb )\right\} \\
R_{\nb} &< \min \left\{ \Delta_1 I(X_{\nb}^\pa ; Y_{\nr}^\pa | X_{\na}^\pa,Q),
            \Delta_2 I(X_{\nr}^\pb ; Y_{\na}^\pb )\right\}\\
R_{\na} + R_{\nb} &<    \Delta_1 I(X_{\na}^\pa , X_{\nb}^\pa ; Y_{\nr}^\pa|Q)
\end{align}
over all joint distributions,
$p^\pa(q)p^\pa(x_\na|q)p^\pa(x_\nb|q)p^\pa(y_\nr|x_\na,x_\nb)p^\pb(x_\nr)p^\pb(y_\na,y_\nb|x_\nr)$
with $|{\cal Q}| \leq 3$ over the restricted alphabet ${\cal X}_\na \times {\cal X}_\nb \times {\cal X}_\nr
\times {\cal Y}_\na \times {\cal Y}_\nb \times {\cal Y}_\nr $.
\thmend
\end{theorem}

\begin{theorem}\label{theorem:TDBC:one}
An achievable rate region of the half-duplex bi-directional relay channel with the decode and forward TDBC protocol is
the closure of the set of all points $(R_\na,R_\nb)$ satisfying
\begin{align}
R_\na &<\min\left\{{\Delta}_1 I(X_\na^\pa;Y_\nr^\pa),{\Delta}_1 I(X_\na^\pa;Y_\nb^\pa) + {\Delta}_3 I(X_\nr^\pc;Y_\nb^\pc)\right\}\\
R_\nb &<\min\left\{{\Delta}_2 I(X_\nb^\pb;Y_\nr^\pb),{\Delta}_2 I(X_\nb^\pb;Y_\na^\pb) + {\Delta}_3 I(X_\nr^\pc;Y_\na^\pc)\right\}
\end{align}
over all joint distributions,
$p^\pa(x_\na)p^\pa(y_\nb,y_\nr|x_\na)p^\pb(x_\nb)p^\pb(y_\na,y_\nr|x_\nb)p^\pc(x_\nr)p^\pc(y_\na,y_\nb|x_\nr)$
over the restricted alphabet ${\cal X}_\na \times {\cal X}_\nb \times {\cal X}_\nr
\times {\cal Y}_\na^2 \times {\cal Y}_\nb^2 \times {\cal Y}_\nr^2 $.
\thmend
\end{theorem}


\section{Protocols}

\label{sec:protocols} We next describe a class of bi-directional
multiple-relay protocols which we term  {\em $(m,t)$ DF MHMR}
(Decode and Forward, Multiple Hop Multiple Relay) protocols, where
$m$ is the number of relays and $t$ is the number of hops. A
protocol is a  series of temporal phases through which
bi-directional communication between nodes $\na$ and $\nb$ is
enabled.  A single protocol may employ different types of relaying schemes, which specify how relays process and forward the received signals. In Section \ref{sec:gaussian} and \ref{sec:regions} we
consider Amplify and Forward relaying in the Gaussian channel and
use the term {\em $(m,t)$ AF MHMR} protocol to denote the protocols described next
 with Amplify and Forward  rather than Decode and
Forward relaying.

When the number of hops is 2, i.e. $t=2$ we re-name the $(m,2)$ MHMR protocol the MABC MHMR protocol. When the number of hops is 3, we re-name the $(m,3)$ MHMR protocol the TDBC MHMR protocol. These names reflect the similarity of the protocols to the previously defined MABC and TDBC protocols \cite{SKim:2007}. We also note that the $(m,3)$ is not a generalization/extension of the $(m,2)$ protocol, and that the  DF MHMR $(m,t)$ DF MHMR protocol for $t>3$ is {\it not} an explicit generalization of the $(m,2)$ and $(m,3)$ protocols, despite the similar notation.

For $t>2$, we define the $(m,t)$ {\em regular} MHMR protocol for $m\mod (t-2)=0$
as the MHMR protocol which has the same number of relays in each
intermediate hop, equal to $m/(t-2)$. For example, the
$(8,6)$ regular MHMR protocol consists of two relays in each of the
four hops.

\subsection{$(m,2)$ MABC and $(m,3)$ TDBC DF MHMR protocols}
If multiple relays are permitted to transmit in a single temporal phase, or hop, the protocols match those of when only a single relay is present \cite{SKim:2007}.  The added complication lies in which subset of  relays will transmit in that phase.  We thus extend the MABC and TDBC protocols previously proposed for the single relay bi-directional channel \cite{SKim:2007} to allow for multiple relays. During the relay transmission phase, each relay lies in one of the four following sets, which partition ${\cal R}$:
\begin{enumerate}
  \item $({\cal A}\cup{\cal B})^c$ : cannot decode $w_\na$ or $w_\nb$
  \item ${\cal A}\setminus {\cal B}$ : decode $w_\na$ only
  \item ${\cal B}\setminus {\cal A}$ : decode $w_\nb$ only
  \item ${\cal A}\cap {\cal B}$ : decode both $w_\na$ and $w_\nb$
\end{enumerate}
Note that the relays in case 1) do not re-transmit any messages as they were not able to decode any,  and the single relay protocol is contained in case 4) (where all relays in ${\cal A}\cap {\cal B}$ may be viewed as a single relay with multiple antennas). With the MHMR protocols, relays in the sets ${\cal A}\setminus {\cal B}$ and ${\cal B}\setminus {\cal A}$ can be used for the relay transmission. The detailed protocol is as follows:
\begin{enumerate}
\item [(a)] Two terminal nodes transmit their own messages.
\begin{itemize}
  \item with the MABC protocol, terminal nodes transmit simultaneously in a single multiple access phase
  \item with the TDBC protocol, terminal nodes transmit in two sequential phases
\end{itemize}

\item [(b)] Relays in ${\cal A}\cap {\cal B}$ generate ${\bf x}_{{\cal A}\cap {\cal B}}(w_\na \oplus w_\nb)$, relays in $ {\cal A}\setminus {\cal B}$ generate ${\bf x}_{{\cal A}\setminus {\cal B}}(w_\na)$ and relays in ${\cal B}\setminus {\cal A}$ generates ${\bf x}_{{\cal B}\setminus {\cal A}}(w_\nb)$ which they simultaneously transmit during the relay transmit phase.
\item [(c)] Node $\na$ receives ${\bf y}_\na$ and decodes $\tilde{w}_\nb$ from the jointly typical sequences $({\bf x}_{{\cal A}\cap {\cal B}},{\bf x}_{{\cal A}\setminus {\cal B}},{\bf x}_{{\cal B}\setminus {\cal A}}, {\bf y}_\na)$. Since $\na$ knows $w_\na$, we can remove ${\bf x}_{{\cal A}\setminus {\cal B}}$ and the total cardinality is bounded by $\lfloor 2^{nR_\nb}\rfloor$. Node $\nb$ similarly decodes $\tilde{w}_\na$.
\end{enumerate}

\begin{figure}[th]
\begin{center}
\epsfig{keepaspectratio = true, width = 7cm, figure =
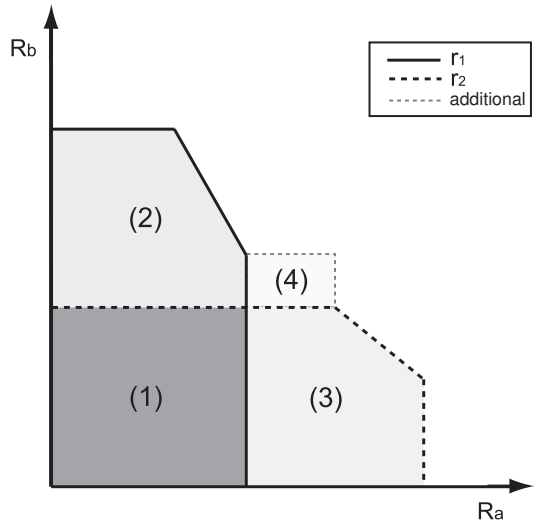}
\end{center}
\caption{Achievable regions of $\nr_1$ and $\nr_2$ for the first
phase (the region in which $\nr_1$ (resp. $\nr_2$) can decode both messages) and the corresponding achievable region (the region in which $w_\na$ (resp. $w_\nb$) is decoded by either $\nr_1$ or $\nr_2$) for the MABC protocol with two relays. } \label{fig:region1}
\end{figure}

Fig. \ref{fig:region1} illustrates the impact of having different subsets ${\cal A}, {\cal B} \subseteq {\cal R}$ in the DF MABC MHMR protocol with two relays.  The labeled rate regions correspond to the different sets ${\cal A}$ and ${\cal B}$ specified in Table \ref{table:multi-relay}. We see that if  ${\cal A} \cap {\cal B} = \emptyset$ then larger rates $R_\na$ and $R_\nb$ may be possible (as in for example region (4)).

\begin{table}
\caption{${\cal A}$ and ${\cal B}$ for each achievable region with ${\cal R} =\{\nr_1,\nr_2\}$}
\label{table:multi-relay}
\centering
\begin{tabular}{c||c|c}
  \hline
  Region & ${\cal A}$ & ${\cal B}$ \\
  \hline
  (1) & $\{\nr_1,\nr_2\}$ & $\{\nr_1,\nr_2\}$ \\
  (2) & $\{\nr_1,\nr_2\}$ & $\{\nr_1\}$ \\
  (3) & $\{\nr_2\}$ & $\{\nr_1,\nr_2\}$ \\
  (4) & $\{\nr_2\}$ & $\{\nr_1\}$ \\
  \hline
\end{tabular}
\end{table}

\subsection{$(m, t)$ DF MHMR protocol}

In this section, we consider a relay network with $m$ relays and $3
< t \leq m+2$. For simplicity, we first describe the $(m,m+2)$ MHMR
protocol: our general protocol with the maximal number of phases.
From the $(m,m+2)$ MHMR protocol the $(m, t)$ for $3<t< m+2$
protocol and corresponding achievable rate regions readily follow.
The multi-hop network may be represented graphically: each node is
represented as a vertex and a directed edge $(s,t)$ exists if node
$t$ can decode $w_\na$ or $w_\nb$ at the end of the transmission of
node $s$. For example,
\begin{align*}
\nr_0 (=\na) \leftrightarrows \nr_1 \leftrightarrows \nr_2
\leftrightarrows \cdots \leftrightarrows \nr_m \leftrightarrows
\nr_{m+1} ( = \nb)
\end{align*}
is one possible graphical representation of our multi-hop network with $m$ relays. A simple na\"{i}ve protocol
for the above example network is : $\nr_0 \rightarrow \nr_1 \rightarrow
\cdots \rightarrow \nr_m \rightarrow \nr_{m+1}$ and then $\nr_{m+1}
\rightarrow \nr_m \rightarrow \cdots \rightarrow \nr_1 \rightarrow
\nr_0$. This  is one possible $(m,2m+2)$ MHMR protocol, which may be spectrally inefficient as the number of phases is large. Intuitively, spectral efficiency may be improved by combining phases through the use of network coding.
In the following, we reduce the number of phases needed from $2m+2$ to $m+2$.
In the $(m,m+2)$ protocol only a single relay transmits during each phase. This is extended to allow for multiple relays transmitting in each phase in Corollary \ref{theorem:DFMH2} of the next section. The protocol may be described in Table \ref{table:algorithm}:

\begin{table}
\caption{[ALGORITHM] - $(m,m+2)$ DF MHMR protocol}
\label{table:algorithm}
\centering
\begin{tabular}{l}
  \hline
  {\em Preparation} \\
  \hline
  $\na$ and $\nb$ divide their respective messages into $B$ sub-messages, one for each \emph{block}. \\Thus, $\na$ has
  the message set $\{w_{\na,(0)},w_{\na,(1)},\cdots,w_{\na,(B-1)}\}$. \\
  Likewise $\nb$ has $\{w_{\nb,(0)},w_{\nb,(1)},\cdots,w_{\nb,(B-1)}\}$.\\
\hline
{\em Initialization}\\
\hline
  01: For $i=0$ to $m-1$\\
  02: ~~~ For $j=0$ to $i$\\
  03: ~~~ ~~~ $\nr_{i-j}$ transmits ${\bf x}_{\nr_{i-j}} (w_{\na,(j)})$\\
  04: ~~~ ~~~ $\nr_{i-j+1}$ decodes $w_{\na,(j)}$ \\
  05: ~~~ end\\
  06: end\\
  \hline
  {\em Main routine}\\
  \hline
  01: For $i=0$ to $B-m-1$\\
  02: ~~~ $\nr_{m+1}$ transmits ${\bf x}_{\nr_{m+1}} (w_{\nb,(i)})$\\
  03: ~~~ $\nr_{m}$ decodes $w_{\nb,(i)}$ and generates ${\bf x}_{\nr_{m}}(w_{\na,(i)} \oplus w_{\nb,(i)})$\\
  04: ~~~ For $j=0$ to $m-1$\\
  05: ~~~ ~~~ $\nr_{m-j}$ transmits ${\bf x}_{\nr_{m-j}}
      (w_{\na,(i+j)} \oplus w_{\nb,(i)})$\\
  06: ~~~ ~~~ $\nr_{m-j-1}$ decodes $w_{\nb,(i)}$ and generates ${\bf x}_{\nr_{m-j-1}}(w_{\na,(i+j+1)} \oplus w_{\nb,(i)})$ \\
  07: ~~~ ~~~ $\nr_{m-j+1}$ decodes $w_{\na,(i+j)}$\\
  08: ~~~  end\\
  09: ~~~ $\nr_{0}$ transmits ${\bf x}_{\nr_{0}} (w_{\na,(m+i)})$\\
  10: ~~~ $\nr_{1}$ decodes $w_{\na,(m+i)}$\\
  11: end\\
  \hline
  {\em Termination}\\
  \hline
  01: For $i=B-m$ to $B-1$\\
  02: ~~~ $\nr_{m+1}$ transmits ${\bf x}_{\nr_{m+1}}(w_{\nb,(i)})$\\
  03: ~~~ $\nr_{m}$ decodes $w_{\nb,(i)}$ and generates ${\bf x}_{\nr_{m}}(w_{\na,(i)} \oplus w_{\nb,(i)})$\\
  04: ~~~ For $j=0$ to $m-1$\\
  05: ~~~ ~~~ $\nr_{m-j}$ transmits ${\bf x}_{\nr_{m-j}}$\\
  06: ~~~ ~~~ $\nr_{m-j-1}$ decodes $w_{\nb,(i)}$ and generates\\
  ~~~ ~~~ ~~~ ~~~ ~~~   $\left\{
        \begin{array}{ll}
           {\bf x}_{\nr_{m-j-1}}
      (w_{\na,(i+j+1)} \oplus w_{\nb,(i)}), & \hbox{if } ~i+j\leq B-2 \\
          {\bf x}_{\nr_{m-j-1}}
      (w_{\nb,(i)}), & \hbox{otherwise}
        \end{array}
      \right.$\\
  07: ~~~ ~~~ $\nr_{m-j+1}$ decodes $w_{\na,(i+j)}$ if $i+j\leq B-1$\\
  08: ~~~ end\\
  09: end\\
  \hline
\end{tabular}
\end{table}

After initialization, relay $\nr_i$ has the following messages from node $\na$: $\{w_{\na,(0)},w_{\na,(1)},\cdots,w_{\na,(m-i)}\}$ $(1\leq i \leq m)$. In other words message $w_{\na,(i)}$  has reached $\nr_{m-i}$ at the end of the initialization. In the main routine, which, when the number of blocks $B\rightarrow \infty$ makes up the majority of this protocol,  $w_{\nb,(i)}$ travels along the path $\nr_{m+1} \rightarrow \nr_m \rightarrow \cdots \rightarrow \nr_1 \rightarrow  \nr_0$ in the $i^{th}$ loop. During the same loop, as the single sub-message from node $\nb$ travels to node $\na$,  the stream of messages from node $\na$ sitting in the each of the relays are all shifted to the right by one through the use of  network coding. Overall then, we require 2 transmissions  from the terminal nodes, and $m$ relay transmissions  to transfer two individual sub-messages. When node $\na$ finishes sending its all sub-messages to $\nr_1$, the termination step starts. The remaining $w_{\na,(i)}$s in the relays and $w_{\nb,(i)}$s in node $\nb$ are processed in this step. The number of transmissions in the main routine depends only on the number of blocks $B$ while the number of transmissions in the initialization and termination steps are a function of the hop size $m$. In the following theorem we formally prove that  by increasing the block size $B$, our algorithm asymptotically results in $m+2$ phases.
\begin{theorem}
The number of phases achieved by the $(m,m+2)$ algorithm with $B$ blocks approaches $m+2$ as the number of blocks $B\rightarrow \infty$.
\thmend
\end{theorem}

\begin{proof}
In the $(m,m+2)$ DF MHMR protocol, the total number of transmissions $N_T(m)$ for $B$
blocks is:
\begin{align}
N_T(m) &= \frac{m(m+1)}{2} + (B-m)(m+2) + m(m+1) \\
&= B(m+2) +\frac{m(m-1)}{2}
\end{align}
Therefore, the number of phases per block is given by
\begin{align}
N_T(m) / B = (m+2) + \frac{m(m-1)}{2B}
\end{align}
As $B\rightarrow \infty$,  $ m+2$ phases result.
\end{proof}

\begin{figure}
\begin{center}
\epsfig{figure=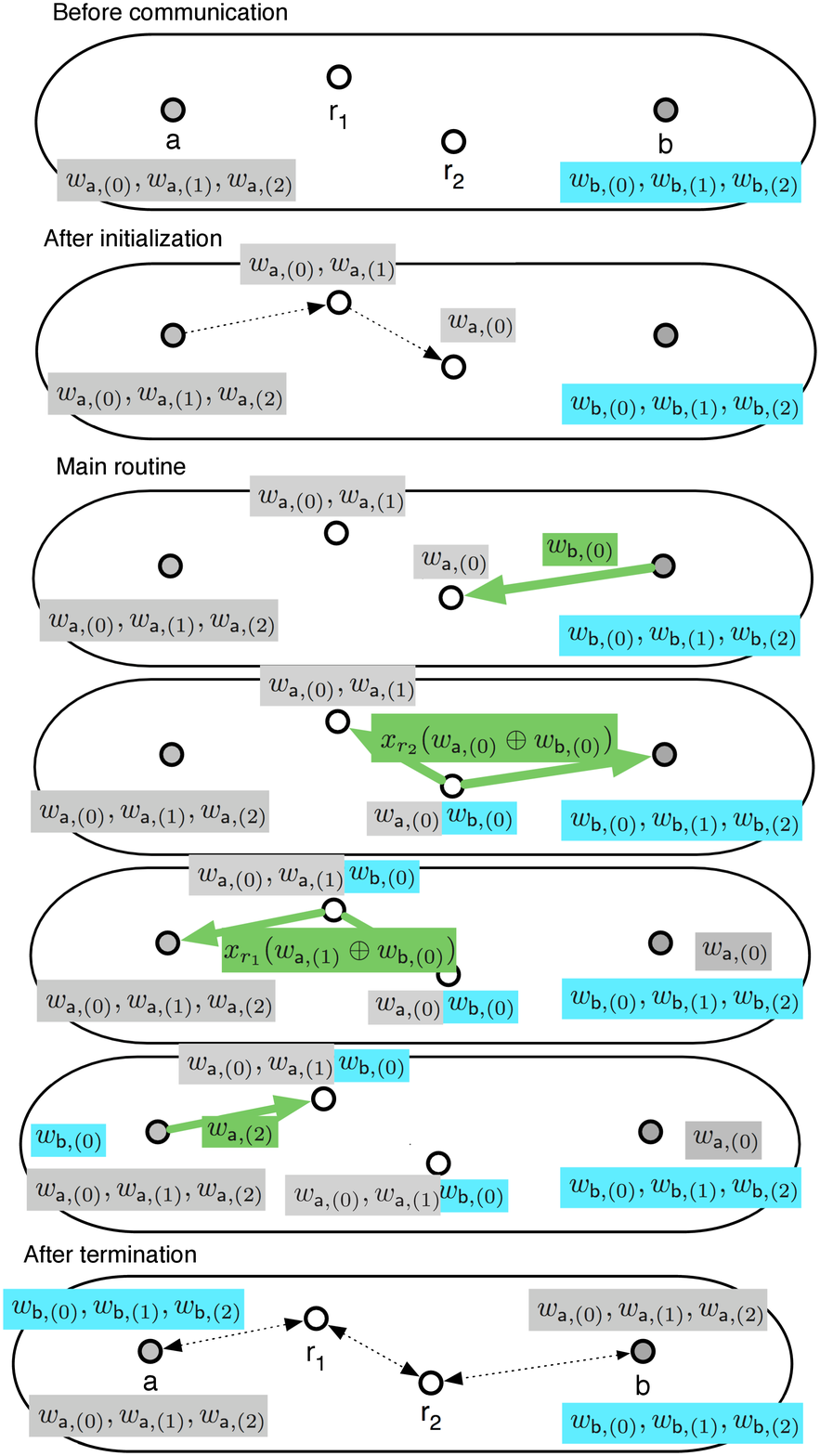, width=10cm}
\end{center}
\caption{Illustration of the $(m,m+2)$ DF MHMR protocol with B=3, m=2. Grey denotes the sub-messages of $w_\na$ at the nodes, blue denotes the sub-messages of $w_\nb$ at the nodes, and green denotes the current transmission. Dotted lines denote the path taken during initialization and termination phases.}
\label{fig:m-m+2}
\end{figure}

A graphical illustration for the case when $B=3$ and $m=2$ is shown in Fig. \ref{fig:m-m+2}.

\section{Achievable rate regions}

\label{sec:bounds}

\subsection{$(m,2)$ DF MABC protocol}
We now derive an achievable rate region for the multi-hop
bi-directional relay channel with $m$ relays and two phases, an
extension of the MABC protocol of  \cite{SKim:2007}  to multiple relays. We note that relays which share the same message set may cooperate in transmitting their messages, as seen in the joint distributions of phase 2.

\begin{theorem}
\label{theorem:MABC_mrop} An achievable region of the half-duplex
bi-directional channel under the $(m,2)$ DF MABC protocol with a given ${\cal A} , {\cal B} \subseteq {\cal R}$ is the
closure of the set of all points $(R_\na,R_\nb)$ satisfying
\begin{align}
R_\na &< \min\big\{{\Delta}_1 I_{\cal{A}\cap \cal{B}}^{\min}(X_\na^\pa;Y_\nr^\pa|X_\nb^\pa,Q),{\Delta}_1 I_{\cal{A}\setminus \cal{B}}^{\min}(X_\na^\pa;Y_\nr^\pa|Q) , {\Delta}_2 I(X_{\cal{A}}^\pb;Y_\nb^\pb|Q)\big\} \label{eq:MABC_mrop_1}\\
R_\nb &< \min\big\{{\Delta}_1 I_{\cal{A}\cap \cal{B}}^{\min}(X_\nb^\pa;Y_\nr^\pa|X_\na^\pa,Q),{\Delta}_1 I_{\cal{B}\setminus \cal{A}}^{\min}(X_\nb^\pa;Y_\nr^\pa|Q) , {\Delta}_2 I(X_{\cal{B}}^\pb;Y_\na^\pb|Q)\big\} \label{eq:MABC_mrop_2}\\
R_\na + R_\nb &< \Delta_1 I_{\cal{A}\cap \cal{B}}^{\min}(X_\na^\pa,
X_\nb^\pa;Y_\nr^\pa|Q) \label{eq:MABC_mrop_3}
\end{align}
over all joint distributions
$p(q)p^\pa(x_{\na}|q)p^\pa(x_{\nb}|q)p^\pa(y_{\cal R}|x_\na,x_\nb)p^\pb(x_{\cal{A}\cap
\cal{B}}|q)p^\pb(x_{\cal{A}\setminus \cal{B}}|q)p^\pb(x_{\cal{B}\setminus \cal{A}}|q)$ $p^\pb(y_\na,y_\nb|x_{\cal{A}\cup
\cal{B}})$ with $|{\cal Q}| \leq 3m +2$ over
the restricted alphabet $\bigotimes_{i=0}^{m+1} ({\cal X}_{\nr_i}\times {\cal Y}_{\nr_i})$. \thmend
\end{theorem}

\begin{proof}
{\em Random code generation: } For simplicity of exposition, we take
$|{\cal Q}| = 1$ and therefore consider distributions
$p^\pa(x_\na)$, $p^\pa(x_\nb)$, $p^\pb(x_{\cal{A}\cap
\cal{B}})$, $p^\pb(x_{\cal{A}\setminus \cal{B}})$ and
$p^\pb(x_{\cal{B}\setminus \cal{A}})$. First we generate
random $(n\cdot \Delta_{1,n})$-length sequences
$\mathbf{x}^{(1)}_\na(w_\na)$ with $w_\na \in {\cal S}_\na$ and
$\mathbf{x}^{(1)}_\nb(w_\nb)$ with $w_\nb \in {\cal S}_\nb$
according to $p^\pa(x_\na)$, $p^\pa(x_\nb)$ to be used in phase 1.
For phase 2, we generate random $(n\cdot\Delta_{2,n})$-length
sequences $\mathbf{x}^{(2)}_{\cal{A}\cap \cal{B}}(w_\nr)$ with
$w_\nr \in \mathbb{Z}_L$ ($L = \max(\lfloor 2^{nR_{\na}} \rfloor,
\lfloor 2^{nR_{\nb}} \rfloor )$),
$\mathbf{x}^{(2)}_{\cal{A}\setminus \cal{B}}(w_\na)$ and
$\mathbf{x}^{(2)}_{\cal{B}\setminus \cal{A}}(w_\nb)$, according to
$p^\pb(x_{\cal{A}\cap \cal{B}})$, $p^\pb(x_{\cal{A}\setminus \cal{B}})$ and $p^\pb(x_{\cal{B}\setminus
\cal{A}})$ respectively.

{\em Encoding: } During phase 1, encoders of node $\na$ and $\nb$ send the codewords ${\bf x}^\pa_\na(w_\na)$ and ${\bf x}^\pa_\nb(w_\nb)$ respectively. Relays in ${\cal A}\cap {\cal B}$ estimate $\hat{w}_\na$ and $\hat{w}_\nb$ after phase 1 using jointly typical decoding, then construct $w_{\nr} =\hat{w}_\na \oplus \hat{w}_\nb$ in ${\mathbb Z}_L$ and send
${\bf x}^\pb_{{\cal A}\cap {\cal B}}(w_\nr)$ during phase 2. Likewise relays in ${\cal A}\setminus {\cal B}$ (resp. ${\cal B}\setminus {\cal A}$) estimate $\hat{w}_\na$ (resp. $\hat{w}_\nb$) after phase 1 and send ${\bf x}^\pb_{{\cal A}\setminus {\cal B}}({\hat w}_\na)$ (resp. ${\bf x}^\pb_{{\cal B}\setminus {\cal A}}({\hat w}_\nb)$).

{\em Decoding: } $\na$ and $\nb$ estimate $\tilde{w}_\nb$ and
$\tilde{w}_\na$ after phase 2 using jointly typical decoding. Since
$w_\nr = w_\na \oplus w_\nb$ and $\na$ knows $w_\na$, node $\na$ can
reduce the cardinality of $w_\nr$ to $\lfloor 2^{nR_{\nb}} \rfloor$.
$\nb$ similarly decodes ${\tilde w}_\na$.

{\em Error analysis: }
\begin{align}
  P[E_{\na,\nb}] & \leq P[E_{\{\na\},{\cal A}\cap {\cal B}}^\pa \cup E_{\{\nb\},{\cal A}\cap {\cal B}}^\pa \cup E_{\{\na\},{\cal A}\setminus {\cal B}}^\pa  \cup E_{{\cal A},\{\nb\}}^\pb]\\
  & \leq P[E_{\{\na\},{\cal A}\cap {\cal B}}^\pa \cup E_{\{\nb\},{\cal A}\cap {\cal B}}^\pa] + P[E_{\{\na\},{\cal A}\setminus {\cal B}}^\pa] + P[E_{{\cal A},\{\nb\}}^\pb | \bar{E}_{\{\na\},{\cal A}\cap {\cal B}}^\pa \cap \bar{E}_{\{\nb\},{\cal A}\cap {\cal B}}^\pa \cap \bar{E}_{\{\na\},{\cal A}\setminus {\cal B}}^\pa]
\end{align}
Following the well-known MAC error analysis from (15.72) in \cite{Cover:2006}:
\begin{align}
P[E_{\{\na\},{\cal A}\cap {\cal B}}^\pa \cup E_{\{\nb\},{\cal A}\cap {\cal B}}^\pa]
&\leq \sum_{\nr \in {\cal A}\cap {\cal B}}  P[E_{\na,\nr}^\pa \cup E_{\nb,\nr}^\pa]\\
&\leq \sum_{\nr \in {\cal A}\cap {\cal B}} \left\{ P[\bar{D}^\pa({\bf x}_\na(w_\na),{\bf x}_\nb(w_\nb),{\bf y}_\nr)] + \right.\nonumber\\
    &~~~~~~~~~~~~~~ 2^{nR_\na} 2^{-n\cdot\Delta_{1,n}(I(X_\na^\pa;Y_\nr^\pa|X_\nb^\pa)-3\epsilon)} +  \nonumber\\
    &~~~~~~~~~~~~~~ 2^{nR_\nb} 2^{-n\cdot\Delta_{1,n}(I(X_\nb^\pa;Y_\nr^\pa|X_\na^\pa)-3\epsilon)} +  \nonumber\\
    &\left.~~~~~~~~~~~~~  2^{n(R_\na+R_\nb)} 2^{-n\cdot\Delta_{1,n}(I(X_\na^\pa,X_\nb^\pa;Y_\nr^\pa)-4\epsilon)} \right\}.\label{eq:mrop1}
\end{align}
  Also,
\begin{align}
 P[E_{\{\na\},{\cal A}\setminus {\cal B}}^\pa] &\leq \sum_{\nr \in {\cal A}\setminus {\cal B}} P[E_{\na,\nr}^\pa]\\
 &\leq \sum_{\nr \in {\cal A}\setminus {\cal B}} P[\bar{D}^\pa({\bf x}_\na(w_\na),{\bf y}_\nr)] + 2^{nR_\na} 2^{-n\cdot\Delta_{1,n}(I(X_\na^\pa;Y_\nr^\pa)-3\epsilon)}, \label{eq:mrop2}
\end{align}
and
\begin{align}
 P[E_{{\cal A},\{\nb\}}^\pb |& \bar{E}_{\{\na\},{\cal A}\cap {\cal B}}^\pa \cap \bar{E}_{\{\nb\},{\cal A}\cap {\cal B}}^\pa \cap \bar{E}_{\{\na\},{\cal A}\setminus {\cal B}}^\pa]\nonumber\\
 &\leq P[\bar{D}^\pb({\bf x}_{{\cal A}\cap{\cal B}}(w_\na\oplus w_\nb),{\bf x}_{{\cal A}\setminus {\cal B}}(w_\na),{\bf y}_\nb)] + 2^{nR_\na} 2^{-n\cdot\Delta_{2,n}(I(X_{\cal A}^\pb;Y_\nb^\pb)-3\epsilon)}. \label{eq:mrop3}
\end{align}

 Since $\epsilon > 0$ is arbitrary, the conditions of Theorem
\ref{theorem:MABC_mrop} and the AEP property will guarantee that the
right hand sides of \eqref{eq:mrop1}, \eqref{eq:mrop2} and \eqref{eq:mrop3}
vanish as $n \rightarrow \infty$. Similarly, $P[E_{\nb,\na}]
\rightarrow 0$ as $n \rightarrow \infty$.
By  Fenchel-Bunt's extension of Carath\'{e}odory
theorem in \cite{Hiriart:2001}, it is sufficient to restrict $|{\cal
Q}| \leq 3m+2$  since $m+1$ inequalities are from \eqref{eq:MABC_mrop_1}, $m+1$ inequalities are from \eqref{eq:MABC_mrop_2} and  at most $m$ inequalities are from \eqref{eq:MABC_mrop_3}.

To apply the time sharing random variable $Q$, we split
\begin{align}
R_\na &= \sum_q p(q) R_{\na,q}\label{eq:mrop4}\\
R_\nb &= \sum_q p(q) R_{\nb,q}\label{eq:mrop5}
\end{align}
Then each $(R_{\na,q},R_{\nb,q})$ pair is in
\begin{align}
R_{\na,q} &< \min\left\{{\Delta}_1 I_{\cal{A}\cap \cal{B}}^{\min}(X_\na^\pa;Y_\nr^\pa|X_\nb^\pa,Q=q),\right.\nonumber\\
&~~~~~~~~~~\left.{\Delta}_1 I_{\cal{A}\setminus \cal{B}}^{\min}(X_\na^\pa;Y_\nr^\pa|Q=q) , {\Delta}_2 I(X_{\cal{A}}^\pb;Y_\nb^\pb|Q=q)\right\}\label{eq:mrop6} \\
R_{\nb,q} &< \min\left\{{\Delta}_1 I_{\cal{A}\cap \cal{B}}^{\min}(X_\nb^\pa;Y_\nr^\pa|X_\na^\pa,Q=q),\right.\nonumber\\
&~~~~~~~~~~\left.{\Delta}_1 I_{\cal{B}\setminus \cal{A}}^{\min}(X_\nb^\pa;Y_\nr^\pa|Q=q) , {\Delta}_2 I(X_{\cal{B}}^\pb;Y_\na^\pb|Q=q)\right\}\label{eq:mrop7}\\
R_{\na,q} + R_{\nb,q} &< \Delta_1 I_{\cal{A}\cap \cal{B}}^{\min}(X_\na^\pa,
X_\nb^\pa;Y_\nr^\pa|Q=q)\label{eq:mrop8}
\end{align}
Applying \eqref{eq:mrop4} and \eqref{eq:mrop5} to \eqref{eq:mrop6}, \eqref{eq:mrop7} and \eqref{eq:mrop8} the general expression of \eqref{eq:MABC_mrop_1}, \eqref{eq:MABC_mrop_2} and \eqref{eq:MABC_mrop_3} is achieved.
\end{proof}

\subsection{$(m,3)$ DF TDBC protocol}
We now derive an achievable rate region for the multi-hop
bi-directional relay channel with $m$ relays and 3 phases, an
extension of the DF TDBC protocol of \cite{SKim:2007}. We note that relays which share the same message set may cooperate in transmitting their messages, as seen in the joint distributions of phase 3.
\begin{theorem}
\label{theorem:TDBC_mrop} An achievable region of the half-duplex
bi-directional channel under  the $(m,3)$ DF TDBC protocol with a given ${\cal A} , {\cal B} \subseteq {\cal R}$ is the
closure of the set of all points $(R_\na,R_\nb)$ satisfying
\begin{align}
R_\na &< \min\big\{{\Delta}_1 I_{\cal{A}}^{\min}(X_\na^\pa;Y_\nr^\pa),{\Delta}_1 I(X_\na^\pa;Y_\nb^\pa) + {\Delta}_3 I(X_{\cal{A}}^\pc;Y_\nb^\pc|Q)\big\}\\
R_\nb &< \min\big\{{\Delta}_2
I_{\cal{B}}^{\min}(X_\nb^\pb;Y_\nr^\pb),{\Delta}_2
I(X_\nb^\pb;Y_\na^\pb) + {\Delta}_3 I(
X_{\cal{B}}^\pc;Y_\na^\pc|Q)\big\}
\end{align}
over all joint distributions
$p^\pa(x_{\na})p^\pa(y_{\cal R},y_\nb|x_\na)p^\pb(x_{\nb})p^\pb(y_{\cal R},y_\na|x_\nb)p^\pc(q)p^\pc(x_{\cal{A}\cap
\cal{B}}|q)p^\pc(x_{\cal{A}\setminus \cal{B}}|q)$ $p^\pc(x_{\cal{B}\setminus \cal{A}}|q)p^\pc(y_\na,y_\nb|x_{\cal{A}\cup
\cal{B}})$ with $|{\cal Q}| \leq 2$ over
the restricted alphabet $\bigotimes_{i=0}^{m+1}({\cal X}_{\nr_i}\times {\cal Y}_{\nr_i})$. \thmend
\end{theorem}
 Theorem \ref{theorem:TDBC_mrop} is proven in a similar manner to
Theorem \ref{theorem:MABC_mrop} in Appendix
\ref{appendix:TDBC_mrop}.
\subsection{$(m,t)$ DF MHMR protocol}

The $(m,2)$ and $(m,3)$ protocols were extensions of previously
derived MABC and TDBC protocols to multiple relays. In this section
we derive the rates achieved by the novel $(m,m+2)$ protocol which does not resemble the MABC and TDBC protocols. We recall that in the $(m,m+2)$ MHMR protocol a single relay transmits in each hop. We
then extend the ideas of the $(m,m+2)$ MHMR protocol to derive
achievable rate regions for general $(m,t)$ protocols with
$3<t<m+2$. Recalling that $\na$ and $\nb$ are denoted as $\nr_0$ and
$\nr_{m+1}$ respectively, our main result lies in the following Theorem.
In the $(m,m+2)$ protocol we assume that the order of relays are predetermined, i.e., $r_1, r_2, \cdots r_m$ are given; how this ordering is done lies outside the scope of this work and is left for future work.  We also note that nodes which are not transmitting in a particular phase may listen to the channel and obtain additional side information about the messages which it may exploit when decoding these messages.
\begin{theorem}
\label{theorem:DFMH} An achievable rate region of the half-duplex
bi-directional multi-hop relay channel under the $(m,m+2)$ DF MHMR
protocol ($m>1)$ is the closure of the set of all points
$(R_\na,R_\nb)$ satisfying
\begin{align}
R_\na &< \min_{1\leq k \leq m+1}\left\{\sum_{i=1}^k {\Delta}_{m+3-i} I(X_{\nr_{i-1}}^{(m+3-i)};Y_{\nr_k}^{(m+3-i)})\right\}\\
R_\nb &< \min_{1\leq k \leq m+1}\left\{\sum_{i=1}^k {\Delta}_i
I(X_{\nr_{m+2-i}}^{(i)};Y_{\nr_{m+1-k}}^{(i)})\right\}
\end{align}
over all joint distributions $\prod_{i=1}^{m+2}
p^{(i)}(x_{\nr_{m+2-i}})p^{(i)}(y_{{\cal R}^* \setminus \{\nr_{m+2-i}\}}|x_{\nr_{m+2-i}}) $ over the
restricted alphabet \\
$\bigotimes_{i=0}^{m+1}({\cal X}_{\nr_i}\times {\cal Y}_{\nr_i}) $.  \thmend
\end{theorem}
 The proof is provided in Appendix \ref{appendix:DFMH}.

The minimization is over the number of hops, and results from the need for a series of relays to decode each message. The summation for a given $k$ represents the accumulated amount of information the node $k$ may use to decode message $w_{\na}$ or $w_{\nb}$.

We can extend Theorem \ref{theorem:DFMH} to allow for multiple
relays in each hop. In order to use network coding, we make the
assumption that each relay is able to decode both $w_\na$ and
$w_\nb$. In each hop or phase then, a subset of the nodes will be
able to decode both messages $w_\na$ and $w_\nb$ and may cooperate
in re-transmitting the obtained messages. We denote this subset of
relays in the $i$-th hop as ${\cal R}_i$.
In the $(m,t)$ protocol we assume that the relay subsets ${\cal R}_i \subset {\cal
R}^*$ are predetermined, i.e., ${\cal R}_0 = \{\na\}, {\cal R}_2, \cdots ,{\cal R}_t =\{\nb\}$ are {\it given} and the protocol does not consider how these subsets are chosen. How to select these relay subsets lies outside the scope of this work and may be seen as a relay selection problem \cite{OB08brru,Vaze:2008, Vaze:2008codeornot, Luo:VTC, Madan:2008, Bletsas:2006}.

\begin{corollary}
\label{theorem:DFMH2} An achievable rate region of the half-duplex
bi-directional channel in the $(m,t)$ DF MHMR protocol for $3<t<m+2$
with a given partition of subsets ${\cal R}_i \subset {\cal
R}$ is the closure of the set of all points $(R_\na,R_\nb)$ satisfying
\begin{align}
R_\na &< \min_{1\leq k \leq t-1} \min_{\nr_k \in {\cal R}_k} \left\{\sum_{i=1}^k {\Delta}_{t+1-i} I(X_{{\cal R} _{i-1}}^{(t+1-i)};Y_{\nr_k}^{(t+1-i)})\right\}\\
R_\nb &< \min_{1\leq k \leq t-1}\min_{\nr_{t-1-k} \in {\cal
R}_{t-1-k}}\left\{\sum_{i=1}^k {\Delta}_i I(X_{{\cal R}
_{t-i}}^{(i)};Y_{\nr_{t-1-k}}^{(i)})\right\}
\end{align}
over all joint distributions $\prod_{i=1}^{t} p^{(i)}(x_{{\cal R}_{t-i}})p^{(i)}(y_{\bar{\cal R}_{t-i}}|x_{{\cal R}_{t-i}}) $ over the
restricted alphabet $\bigotimes_{i=0}^{m+1}({\cal X}_{\nr_i}\times {\cal Y}_{\nr_i})$.  \thmend
\end{corollary}
The proof of Corollary \ref{theorem:DFMH2} follows the same
argument as the proof of Theorem \ref{theorem:DFMH}.


\section{Outer Bounds}

\label{sec:outer_bounds}

In this section we derive outer bounds for each MHMR protocol using the following
cut-set bound lemma \cite{SKim:2007}. Again, given subsets $S,T
\subseteq {\cal M} = \{1,2,\cdots,m\}$, and $\bar{S}\eqdef {\cal M}\backslash S$, we define $W_{S,T} \eqdef
\{W_{i,j} | i\in S , j\in T\}$ and $R_{S,T} = \lim_{n\rightarrow
\infty }\frac1n H(W_{S,T})$.

\begin{lemma}
\label{lemma:out} If in some network the information rates
$\{R_{i,j}\}$ are achievable for a protocol $\prot$ with relative
durations $\{\Delta_\ell\}$, then for every $\epsilon>0$ and all $S
\subset  {\cal M}$
\begin{align}
 R_{S,\bar{S}} \leq
   \sum_\ell \Delta_\ell I(X^{(\ell)}_{S};Y^{(\ell)}_{\bar{S}}|X^{(\ell)}_{\bar{S}},Q)+\epsilon,
\end{align}
for a family of conditional distributions $p^{(\ell)}(x_1, x_2,
\ldots, x_m|q)$ and a discrete time-sharing random variable $Q$ with
distribution $p(q)$. Furthermore, each $p^{(\ell)}(x_1, x_2, \ldots,
x_m|q)p(q)$ must satisfy the constraints of phase $\ell$ of protocol
$\prot$. \thmend
\end{lemma}
We next state the outer bounds, which will be numerically evaluated and discussed in the following sections.

\subsection{$(m,2)$ MABC protocol}

\begin{theorem}
\label{theorem:MABC:out} (Outer bound) The capacity region of the
half-duplex bi-directional relay channel with the $(m,2)$ MABC
protocol is outer bounded by the set of rate pairs $(R_{\na}, R_{\nb})$ satisfying
\begin{align}
R_\na &\leq \min_{S_R} \big\{\Delta_1 I(X_\na^\pa;Y_{\bar{S}_R}^\pa
|
X_\nb^\pa,Q) + \Delta_2 I(X_{S_R}^\pb;Y_\nb^\pb|X_{\bar{S}_R}^\pb)\big\} \label{eq:MABC:out:1}\\
R_\nb &\leq \min_{S_R} \big\{\Delta_1 I(X_\nb^\pa;Y_{\bar{S}_R}^\pa
| X_\na^\pa,Q) + \Delta_2
I(X_{S_R}^\pb;Y_\na^\pb|X_{\bar{S}_R}^\pb)\big\}\label{eq:MABC:out:2}
\end{align}
for all choices of the joint distribution
$p^\pa(q)p^\pa(x_{\na}|q)p^\pa(x_{\nb}|q)p^\pa(y_{\cal R}|x_\na,x_\nb)p^\pb(x_{\cal R})p^{\pb}(y_\na,y_\nb|x_{\cal R})$ with $|{\cal
Q}| \leq 2^{m+1}$ over the restricted alphabet $\bigotimes_{i=0}^{m+1}
({\cal X}_{\nr_i}\times {\cal Y}_{\nr_i})$ for all possible $S_R \subseteq {\cal R}$. \thmend
\end{theorem}

\begin{proof}
We use Lemma \ref{lemma:out} to prove the
Theorem \ref{theorem:MABC:out}. For every $S_R\subseteq {\cal R}$,
there exist 4 types of cut-sets such that $S_1 =\{\na\} \cup S_R$, $S_2
=\{\nb\} \cup S_R$, $S_3 = \{\na,\nb\}\cup S_R$ and $S_4 = S_R$, as
well as two rates $R_\na$ and $R_\nb$. Also, in the MABC
protocol,
\begin{align}
 Y_\na^\pa &= Y_\nb^\pa = X_{\cal R}^\pa = \varnothing \label{MABC:out:1} \\
 X_\na^\pb &= X_\nb^\pb = Y_{\cal R}^\pb = \varnothing. \label{MABC:out:2}
\end{align}

Thus, the corresponding outer bounds for a given subset $S_R$ are:
\begin{align}
S_1 &: R_\na \leq \Delta_1 I(X_\na^\pa; Y_{\bar{S}_R}^\pa
|X_\nb^\pa,Q)+\Delta_2 I(X_{S_R}^\pb;Y_\nb^\pb|X_{\bar{S}_R}^\pb)+ \epsilon, \label{MABC:out:3}\\
S_2 &: R_\nb \leq \Delta_1 I(X_\nb^\pa; Y_{\bar{S}_R}^\pa
|X_\na^\pa,Q)+\Delta_2
I(X_{S_R}^\pb;Y_\na^\pb|X_{\bar{S}_R}^\pb)+\epsilon,
\label{MABC:out:4}
\end{align}
where the cut sets $S_3$ and $S_4$ yield no constraints.
Since $\epsilon > 0$ is arbitrary, \eqref{MABC:out:3},
\eqref{MABC:out:4} and the fact that the half-duplex nature of the
channel constrains $X_\na^\pa$ to be conditionally independent of
$X_\nb^\pa$ given $Q$ yields Theorem~\ref{theorem:MABC:out}. By
Fenchel-Bunt's extension of the Carath\'{e}odory theorem in
\cite{Hiriart:2001}, it is sufficient to restrict $|{\cal Q}| \leq
2^{m+1}$ since $2^m$ inequalities are from \eqref{eq:MABC:out:1} and $2^m$ inequalities are from \eqref{eq:MABC:out:2}.
\end{proof}

\subsection{$(m,3)$ TDBC protocol}
\begin{theorem}
\label{theorem:TDBC:out} (Outer bound) The capacity region of the
half-duplex bi-directional relay channel with the $(m,3)$ TDBC
protocol is outer bounded  by the set of rate pairs $(R_{\na}, R_{\nb})$ satisfying
\begin{align}
R_\na &\leq \min_{S_R} \big\{\Delta_1 I(X_\na^\pa;Y_{\bar{S}_R}^\pa
,Y_\nb^\pa) + \Delta_3 I(X_{S_R}^\pc;Y_\nb^\pc|X_{\bar{S}_R}^\pc)\big\}\\
R_\nb &\leq \min_{S_R} \big\{\Delta_2 I(X_\nb^\pb;Y_{\bar{S}_R}^\pb
, Y_\na^\pb) + \Delta_3
I(X_{S_R}^\pc;Y_\na^\pc|X_{\bar{S}_R}^\pc)\big\}
\end{align}
for all choices of the joint distribution
$p^\pa(x_{\na})p^\pa(y_\nb,y_{\cal R}|x_\na)p^\pb(x_{\nb})p^\pb(y_\na,y_{\cal R}|x_\nb)p^\pc(x_{\cal R})p^\pc(y_\na,y_\nb|x_{\cal R})$ over the restricted alphabet $\bigotimes_{i=0}^{m+1}
{\cal X}_{\nr_i}$ for all possible $S_R \subseteq {\cal R}$. \thmend
\end{theorem}
Theorem \ref{theorem:TDBC:out} is proven in a similar manner to
Theorem \ref{theorem:MABC:out} in Appendix \ref{appendix:TDBC:out}.

\subsection{$(m,t)$ MHMR protocol}

In the $(m,m+2)$ protocol we assume that the order of the relays is predetermined, i.e., $r_1, r_2, \cdots r_m$ are given; how this ordering is done lies outside the scope of this work and is left for future work.
\begin{theorem}
\label{theorem:MHMR:1} (Outer bound) The capacity region of the
half-duplex bi-directional multi-hop relay channel under the
$(m,m+2)$ MHMR protocol ($m>1)$ is outer bounded  by the set of rate pairs $(R_{\na}, R_{\nb})$ satisfying
\begin{align}
R_\na &\leq \min_{S_R}\left\{\sum_{\nr_i\in S_R\cup \{\na\}} {\Delta}_{m+2-i} I(X_{\nr_{i}}^{(m+2-i)};Y_{{\bar S}_R}^{(m+2-i)},Y_\nb^{(m+2-i)})\right\}\\
R_\nb &\leq \min_{S_R}\left\{\sum_{\nr_i\in S_R\cup \{\nb\}}
{\Delta}_{m+2-i} I(X_{\nr_{i}}^{(m+2-i)};Y_{{\bar
S}_R}^{(m+2-i)},Y_\na^{(m+2-i)})\right\}
\end{align}
for all choices of the joint distribution $\prod_{i=1}^{m+2}
p^{(i)}(x_{\nr_{m+2-i}})p^{(i)}(y_{{\cal R}^* \setminus \{\nr_{m+2-i}\}}|x_{\nr_{m+2-i}}) $ over the
restricted alphabet $\bigotimes_{i=0}^{m+1}({\cal X}_{\nr_i}\times {\cal Y}_{\nr_i})$ for all
possible $S_R \subseteq {\cal R}$.
 \thmend
\end{theorem}
The proof of Theorem \ref{theorem:MHMR:1} follows the same argument
as the proofs of Theorem \ref{theorem:MABC:out} and Theorem
\ref{theorem:TDBC:out}.

In the $(m,t)$ protocol we assume that the relay subsets ${\cal R}_i \subset {\cal
R}^*$ are predetermined, i.e., ${\cal R}_0 = \{\na\}, {\cal R}_2, \cdots ,{\cal R}_t =\{\nb\}$ are {\it given} and the protocol does not consider how these subsets are chosen. How to select these relay subsets lies outside the scope of this work and may be seen as a relay selection problem \cite{OB08brru,Vaze:2008, Vaze:2008codeornot, Luo:VTC, Madan:2008, Bletsas:2006}.
\begin{corollary}
\label{theorem:MHMR:2} (Outer bound) The capacity region of the
half-duplex bi-directional channel in the $(m,t)$ MHMR protocol for
$3<t<m+2$ is outer bounded by the set of rate pairs $(R_{\na}, R_{\nb})$ satisfying
\begin{align}
R_\na &\leq \min_{S_R} \left\{\sum_{i=0}^{t-1} {\Delta}_{t-i} I(X_{{\cal R}_{i}\cap (S_R\cup \{\na\})}^{(t-i)};Y_{{\bar S}_R\setminus {\cal R}_i}^{(t-i)},Y_\nb^{(t-i)}|X_{{\bar S}_R \cap {\cal R}_i}^{(t-i)})\right\}\\
R_\nb &\leq \min_{S_R} \left\{\sum_{i=0}^{t-1} {\Delta}_{t-i}
I(X_{{\cal R}_{i}\cap (S_R\cup \{\nb\})}^{(t-i)};Y_{{\bar
S}_R\setminus {\cal R}_i}^{(t-i)},Y_\na^{(t-i)}|X_{{\bar S}_R \cap
{\cal R}_i}^{(t-i)})\right\}
\end{align}
over all joint distributions $\prod_{i=1}^{t} p^{(i)}(x_{{\cal R}_{t-i}})p^{(i)}(y_{\bar{\cal R}_{t-i}}|x_{{\cal R}_{t-i}}) $ over the
restricted alphabet $\bigotimes_{i=0}^{m+1}({\cal X}_{\nr_i}\times {\cal Y}_{\nr_i})$ for all
possible $S_R \subseteq {\cal R}$. \thmend
\end{corollary}
The proof of Corollary \ref{theorem:MHMR:2} follows the same
argument as the proofs of Theorem \ref{theorem:MABC:out} and Theorem
\ref{theorem:TDBC:out}.


\section{The Gaussian Relay Network}

\label{sec:gaussian}

In this section, we apply the bounds obtained in the previous section to a Gaussian relay network.
 Since strong typicality does not apply to
continuous random variables, the achievable rate regions from the theorems in the previous section do
not directly apply to continuous domains. However, for the Gaussian input distributions and additive
Gaussian noise which we will assume in the following, the Markov lemma of \cite{Oohama:1997}, which generalizes
the Markov lemma to the continuous domains, ensures that the achievable rate regions in the previous
section hold in the Gaussian case.

We assume that there are two terminal nodes $\na$ and $\nb$, and $m$ relays $\nr_1,\nr_2,\cdots, \nr_m$. Also, for convenience of analysis, we denote $\na$ as $\nr_0$ and $\nb$ as $\nr_{m+1}$. The corresponding mathematical channel model is, for each channel use $k$ :
\begin{align}
{\bf Y}[k] = {\bf H}{\bf X}[k] + {\bf Z}[k]
\end{align}
where,
\begin{align}
{\bf Y}[k] = \left[
               \begin{array}{c}
                 Y_{\nr_0}[k] \\
                 Y_{\nr_1}[k] \\
                 {\vdots} \\
                 Y_{\nr_{m+1}}[k] \\
               \end{array}
             \right] ,~~~
{\bf X}[k] = \left[
               \begin{array}{c}
                 X_{\nr_0}[k] \\
                 X_{\nr_1}[k] \\
                 {\vdots} \\
                 X_{\nr_{m+1}}[k] \\
               \end{array}
             \right] ,~~~
{\bf Z}[k] = \left[
               \begin{array}{c}
                 Z_{\nr_0}[k] \\
                 Z_{\nr_1}[k] \\
                 {\vdots} \\
                 Z_{\nr_{m+1}}[k] \\
               \end{array}
             \right]
\end{align}
and
\begin{align}
{\bf H} = \left[
            \begin{array}{cccc}
              0 & h_{\nr_1,\nr_0} & \cdots & h_{\nr_{m+1},\nr_0} \\
              h_{\nr_0,\nr_1} & 0 & \cdots & h_{\nr_{m+1},\nr_1} \\
              \vdots & \ddots & \vdots & \vdots \\
              h_{\nr_0,\nr_{m+1}} & h_{\nr_1,\nr_{m+1}} & \cdots & 0 \\
            \end{array}
          \right]
\end{align}
where ${\bf Y}[k]$, ${\bf X}[k]$ and ${\bf Z}[k]$ are in $(\complex^*)^{(m+2) \times 1} = (\complex \cup \{\varnothing\})^{(m+2)\times 1}$, and ${\bf H} \in \complex^{(m+2)\times(m+2)}$. In phase $\ell$, if node ${\nr_i}$ is in transmission mode $X_{\nr_i}[k]$ follows the input distribution $X_{\nr_i}^{(\ell)} \sim {\cal N}(0,P_{\nr_i})$. Otherwise, $X_{\nr_i}[k]= \varnothing$, which means that the input symbol does not exist in the above mathematical channel model.

In each phase, the total transmit power is bounded by $P$, i.e.
$\sum_{\nr\in {\cal R}_{\ell}} E[X_\nr^2] \leq P$ for all $\ell$, where ${\cal R}_{\ell}$ is the set of nodes which transmit
during phase $\ell$. While ideally the per-phase power of $P$ could
be distributed amongst the nodes in ${\cal R}_{\ell}$ arbitrarily,
as a first step, we allocate equal power $P/|{\cal R}_\ell|$ for
each relay in ${\cal R}_\ell$. Equal power allocation between
participating nodes may also be simpler to implement. We will later investigate the gain achieved by allowing for arbitrary power allocations.

In each phase, we also allow for cooperation between relays which
have the same messages. For example, in the $(m,2)$ DF MABC
protocol, we have three different subsets of relays in phase 2:
${\cal A}\cap{\cal B}$, ${\cal A}\setminus{\cal B}$ and ${\cal
B}\setminus{\cal A}$. We first allocate equal power $P/|{\cal
A}\cup{\cal B}|$ to each relay in ${\cal A}\cup{\cal B}$ and then
allow cooperation in each subset which has the same messages. For
convenience of analysis we denote $P_{{\cal R}_\ell}$ as the total
power of relays in ${\cal R}_\ell$. $h_{i,j}$ is the effective
channel gain between transmitter $i$ and receiver $j$. We assume the
channel is reciprocal ($h_{i,j} = h_{j,i}$) and that each node is
fully aware of the channel gains, i.e., full CSI. The noise at all
receivers is independent, of unit power, additive, white Gaussian,
complex and circularly symmetric. For convenience of analysis, we
also define the function $C(x) \eqdef \log_2(1+x)$.

\subsection{Amplify and Forward}
As a comparison point for the DF MHMR protocols,  we derive an achievable region of the same temporal protocols in which the relays use a simple amplify and forward relaying scheme rather than a decode and forward scheme. ``Simple'' means that there is no power optimization in each phase, i.e. each node during phase $\ell$ has equal transmit power $P/|{\cal R}_{\ell}|$. Also, in the amplify and forward scheme, all phase durations are equal since relaying is performed on a symbol by symbol basis. Thus, $\Delta_{\ell} = \frac1t$, where $t$ is the number of phases and $\ell \in [1,t]$. Furthermore, relay $\nr$ scales the received symbol by $\sqrt{\frac{P_\nr}{P_{y_\nr}}}$ to meet the transmit power constraint.
We now state the achievable rate regions for the analogous MHMR protocols with AF relaying.

\begin{itemize}
\item {\em $(m,2)$ AF MABC Protocol}
\begin{align}
R_\na &< \frac12 C\left( \frac{\frac{P}{2}\left(\sum_{i=1}^m \sqrt{|h_{\nb,\nr_i}|^2 |h_{\na,\nr_i}|^2 {\tilde P}_{\nr_i}}\right)^2}{\sum_{i=1}^m |h_{\nb,\nr_i}|^2 {\tilde P}_{\nr_i} + 1}\right) \label{eq:afmabc:1}\\
R_\nb &< \frac12 C\left( \frac{\frac{P}{2}\left(\sum_{i=1}^m
\sqrt{|h_{\nb,\nr_i}|^2 |h_{\na,\nr_i}|^2  {\tilde
P}_{\nr_i}}\right)^2}{\sum_{i=1}^m |h_{\na,\nr_i}|^2 {\tilde
P}_{\nr_i} + 1}\right)\label{eq:afmabc:2}
\end{align}
where ${\tilde P}_{\nr_i} = \frac{\frac{P}{m}}{\frac{P}{2} (|h_{\nr_i,\nb}|^2+|h_{\nr_i,\na}|^2)+1}$.\\
\item {\em $(m,3)$ AF TDBC Protocol}
\begin{align}
R_\na &< \frac13 C\left(|h_{\na,\nb}|^2 P +  \frac{P\left(\sum_{i=1}^m \sqrt{|h_{\nb,\nr_i}|^2 |h_{\na,\nr_i}|^2 {\tilde P}_{\nr_i}}\right)^2}{2 \sum_{i=1}^m |h_{\nb,\nr_i}|^2 {\tilde P}_{\nr_i} + 1}\right)\\
R_\nb &< \frac13 C\left(|h_{\na,\nb}|^2 P +
\frac{P\left(\sum_{i=1}^m \sqrt{|h_{\nb,\nr_i}|^2 |h_{\na,\nr_i}|^2
{\tilde P}_{\nr_i}}\right)^2}{2 \sum_{i=1}^m |h_{\na,\nr_i}|^2
{\tilde P}_{\nr_i} + 1}\right)
\end{align}
where ${\tilde P}_{\nr_i} = \frac{\frac{P}{m}}{P(|h_{\nr_i,\nb}|^2+|h_{\nr_i,\na}|^2)+2}$.\\
\item {\em $(m,m+2)$ AF MHMR Protocol}\\
We now consider the  $(m,m+2)$ MHMR protocol  with Amplify and Forward relaying. The temporal protocol consists of the same phases: {\em Initialization}, {\em Main routine}, and {\em Termination}, but  we replace Decode and Forward relaying with Amplify and Forward relaying. In the main routine, we schedule the transmissions as $\nr_{m+1}\rightarrow \nr_m\rightarrow \cdots \rightarrow \nr_0$. To simplify the analysis,  we assume node $\nr_i$ which receives $Y_{\nr_i}^{(\ell)}$ during phase $\ell$  constructs its transmission $X_{\nr_i}^{(m-i+2)}$ as a function of $Y^{(m-i+1)}_{\nr_i}$ and $Y^{(m-i+3)}_{\nr_i}$. That is, it considers only the received symbols from its neighboring nodes  $\nr_{i+1}$ and $\nr_{i-1}$, where $i\in [1,m]$ rather than all previously heard transmissions providing a simpler but smaller achievable rate region. The corresponding channel model is then
\begin{align}
Y_{\nr_i}^{(m-i+1)} &= h_{\nr_{i+1},\nr_i} X_{\nr_{i+1}}^{(m-i+1)} + Z_{\nr_i}^{(m-i+1)} \label{eq:af:1}\\
Y_{\nr_i}^{(m-i+3)} &= h_{\nr_{i-1},\nr_i} X_{\nr_{i-1}}^{(m-i+3)} + Z_{\nr_i}^{(m-i+3)}. \label{eq:af:2}
\end{align}
We construct the channel input symbol $X_{\nr_i}^{(m-i+2)}$  as
\begin{align}
X_{\nr_i}^{(m-i+2)} = \sqrt{{\tilde P}_{\nr_i}} \left(\tilde{h}_{\na,\nr_i} X_\na^{(m+2)} + {\tilde h}_{\nb,\nr_i} X_\nb^\pa + {\tilde Z}_{\na,\nr_i} + {\tilde Z}_{\nb,\nr_i}\right), \label{eq:af:3}
\end{align}
where $\tilde{h}_{\na,\nr_i}$ is the effective channel gain from $\na$ to $\nr_i$ that captures the channel $\na \rightarrow (\nr_1,\nr_2,\cdots,\nr_{i-1})\rightarrow \nr_i$ and ${\tilde h}_{\nb,\nr_i}$ is the effective channel gain from $\nb$ to $\nr_i$ that captures $\nb \rightarrow (\nr_{m},\nr_{m-1},\cdots,\nr_{i+1})\rightarrow \nr_i$ where ${\tilde Z}_{\na,\nr_i}$ and ${\tilde Z}_{\nb,\nr_i}\sim {\cal N} (0,1)$ and ${\tilde P}_{\nr_i} = \frac{P}{P(|{\tilde h}_{\na,\nr_i}|^2 + |{\tilde h}_{\nb,\nr_i}|^2)+2}$. We apply \eqref{eq:af:3} to $X_{\nr_i}^{(m-i+1)}$ and $X_{\nr_i}^{(m-i+3)}$ in \eqref{eq:af:1} and \eqref{eq:af:2} to obtain:
\begin{align}
Y_{\nr_i}^{(m-i+1)} &= h_{\nr_{i+1},\nr_i} \sqrt{{\tilde P}_{\nr_{i+1}}} (\tilde{h}_{\na,\nr_{i+1}} X_\na^{(m+2)} + {\tilde h}_{\nb,\nr_{i+1}} X_\nb^\pa + {\tilde Z}_{\na,\nr_{i+1}} + {\tilde Z}_{\nb,\nr_{i+1}}) + Z_{\nr_i}^{(m-i+1)} \label{eq:af:4}\\
Y_{\nr_i}^{(m-i+3)} &= h_{\nr_{i-1},\nr_i} \sqrt{{\tilde P}_{\nr_{i-1}}} (\tilde{h}_{\na,\nr_{i-1}} X_\na^{(m+2)} + {\tilde h}_{\nb,\nr_{i-1}} X_\nb^\pa + {\tilde Z}_{\na,\nr_{i-1}} + {\tilde Z}_{\nb,\nr_{i-1}}) + Z_{\nr_i}^{(m-i+3)}. \label{eq:af:5}
\end{align}
In \eqref{eq:af:4}, since $X_\na$ flows from $\nr_i$ to $\nr_{i+1}$, $\nr_i$ knows $X_\na^{(m+2)}$ when it receives $Y_{\nr_i}^{(m-i+1)}$. Thus, $\nr_i$ can eliminate $X_\na^{(m+2)}$ from $Y_{\nr_i}^{(m-i+1)}$. By the same reasoning,  $\nr_i$ can eliminate $X_\nb^{(1)}$ from $Y_{\nr_i}^{(m-i+2)}$. After elimination and normalization, the modified ${\tilde Y}_{\nr_i}^{(m-i+1)}$ and ${\tilde Y}_{\nr_i}^{(m-i+3)}$ are given by:
\begin{align}
{\tilde Y}_{\nr_i}^{(m-i+1)} & = \sqrt{\left(\frac{|h_{\nr_{i+1},\nr_i}|^2|{\tilde h}_{\nb,\nr_{i+1}}|^2 {\tilde P}_{\nr_{i+1}} P }{2|h_{\nr_{i+1},\nr_i}|^2 {\tilde P}_{\nr_{i+1}} + 1}\right)}\cdot X_\nb^\pa + {\tilde Z}_{\nb,\nr_i}\label{eq:af:6}\\
& = {\tilde h}_{\nb,\nr_i} X_\nb^\pa + {\tilde Z}_{\nb,\nr_i}\label{eq:af:7}\\
{\tilde Y}_{\nr_i}^{(m-i+3)} & = \sqrt{\left(\frac{|h_{\nr_{i-1},\nr_i}|^2|{\tilde h}_{\na,\nr_{i-1}}|^2 {\tilde P}_{\nr_{i-1}} P }{2|h_{\nr_{i-1},\nr_i}|^2 {\tilde P}_{\nr_{i-1}} + 1}\right)}\cdot X_\na^{(m+2)} + {\tilde Z}_{\na,\nr_i}\label{eq:af:8}\\
& = {\tilde h}_{\na,\nr_i} X_\na^{(m+2)} + {\tilde Z}_{\na,\nr_i}. \label{eq:af:9}
\end{align}
We obtain $X_{\nr_i}^{(m-i+2)}$ by adding the two terms ${\tilde Y}_{\nr_i}^{(m-i+1)}$ and ${\tilde Y}_{\nr_i}^{(m-i+3)}$ and scaling it all by $\sqrt{{\tilde P}_{\nr_i}}$, as
\begin{align}
X_{\nr_i}^{(m-i+2)} = \sqrt{{\tilde P}_{\nr_i}} ({\tilde Y}_{\nr_i}^{(m-i+1)} + {\tilde Y}_{\nr_i}^{(m-i+3)}).
\end{align}
From \eqref{eq:af:6} - \eqref{eq:af:9}, we derive the recurrence relations for $\{|{\tilde h}_{\nb,\nr_i}|^2\}$ and $\{|{\tilde h}_{\na,\nr_i}|^2\}$ for $2\leq i\leq m-1$ as
\begin{align}
|{\tilde h}_{\nb,\nr_i}|^2 &= \frac{P^2 |h_{\nr_{i+1},\nr_i}|^2 |{\tilde h}_{\nb,\nr_{i+1}}|^2}{2P|{\tilde h}_{\nb,\nr_{i+1}}|^2+P(|{\tilde h}_{\na,\nr_{i+1}}|^2+|{\tilde h}_{\nb,\nr_{i+1}}|^2) + 2}\\
|{\tilde h}_{\na,\nr_i}|^2 &= \frac{P^2 |h_{\nr_{i-1},\nr_i}|^2 |{\tilde h}_{\na,\nr_{i-1}}|^2}{2P|{\tilde h}_{\na,\nr_{i-1}}|^2+P(|{\tilde h}_{\nb,\nr_{i-1}}|^2+|{\tilde h}_{\na,\nr_{i-1}}|^2) + 2},
\end{align}
where $|{\tilde h}_{\nb,\nr_m}|^2 = |h_{\nb,\nr_m}|^2$ and $|{\tilde h}_{\na,\nr_1}|^2 = |h_{\na,\nr_1}|^2$. Then an achievable rate region is given by:
\begin{align}
R_\na &< \frac{1}{m+2} C\left(|h_{\na,\nb}|^2P + \sum_{i=1}^m \frac{|h_{\nr_i,\nb}|^2 |\tilde{h}_{\na,\nr_i}|^2 \tilde{P}_{\nr_i} P }{2|h_{\nr_i,\nb}|^2\tilde{P}_{\nr_i} + 1}\right)\\
R_\nb &< \frac{1}{m+2} C\left(|h_{\na,\nb}|^2P + \sum_{i=1}^m
\frac{|h_{\nr_i,\na}|^2 |\tilde{h}_{\nb,\nr_i}|^2 \tilde{P}_{\nr_i}
P }{2|h_{\nr_i,\na}|^2\tilde{P}_{\nr_i} + 1}\right).
\end{align}
\end{itemize}

\subsection{Decode and Forward}
We can likewise obtain the achievable rate regions from Theorems
\ref{theorem:MABC_mrop}, \ref{theorem:TDBC_mrop}, \ref{theorem:DFMH}
and Corollary \ref{theorem:DFMH2} in Gaussian noise, under the same
power allocation assumptions as above as:
\begin{itemize}
\item {\em $(m,2)$ DF MABC Protocol}

The achievable rate region of the  $(m,2)$ DF MABC Protocol is the
union over all $\Delta_1+\Delta_2=1, \Delta_1, \Delta_2\geq 0$ of
\begin{align}
R_\na &< \min\left\{\min_{\nr_i \in {\cal A}\cap{\cal B}} \Delta_1 C\left(\frac{P}{2}|h_{\na,\nr_i}|^2\right), \min_{\nr_j \in {\cal A}\setminus{\cal B}} \Delta_1 C\left(\frac{P|h_{\na,\nr_j}|^2}{P|h_{\nb,\nr_j}|^2+2} \right)  \right\} \label{eq:dfmabc:1}\\
R_\na &<  \Delta_2 C\left(\sum_{\nr_i \in {\cal A}\cap {\cal B}} |h_{\nr_i,\nb}|^2 P_{{\cal A}\cap {\cal B}} + \sum_{\nr_j\in {\cal A}\setminus{\cal B}} |h_{\nr_j,\nb}|^2 P_{{\cal A}\setminus{\cal B}}  \right) \label{eq:dfmabc:2}\\
R_\nb &< \min\left\{\min_{\nr_i \in {\cal A}\cap{\cal B}} \Delta_1 C\left(\frac{P}{2}|h_{\nb,\nr_i}|^2\right), \min_{\nr_j \in {\cal B}\setminus{\cal A}} \Delta_1 C\left(\frac{P|h_{\nb,\nr_j}|^2}{P|h_{\na,\nr_j}|^2+2} \right)  \right\}\label{eq:dfmabc:3}\\
R_\nb &<  \Delta_2 C\left(\sum_{\nr_i \in {\cal A}\cap {\cal B}} |h_{\nr_i,\na}|^2 P_{{\cal A}\cap {\cal B}} + \sum_{\nr_j\in {\cal B}\setminus{\cal A}} |h_{\nr_j,\na}|^2 P_{{\cal B}\setminus{\cal A}}  \right) \label{eq:dfmabc:4}\\
R_\na + R_\nb &\leq \min_{\nr_i \in {\cal A}\cap{\cal B}} \Delta_1 C\left(\frac{P}{2}(|h_{\na,\nr_i}|^2 + |h_{\nb,\nr_i}|^2)\right)\label{eq:dfmabc:5}
\end{align}
where $P_{{\cal A}\cap{\cal B}} = \frac{|{\cal A}\cap{\cal B}|}{|{\cal A}\cup{\cal B}|}P$, $P_{{\cal A}\setminus{\cal B}} = \frac{|{\cal A}\setminus{\cal B}|}{|{\cal A}\cup{\cal B}|}P$, and $P_{{\cal B}\setminus{\cal A}} =\frac{|{\cal B}\setminus{\cal A}|}{|{\cal A}\cup{\cal B}|}P$ over all ${\cal A, B}\subseteq {\cal R}$.\\

\item {\em $(m,3)$ DF TDBC Protocol}

The achievable rate region of the  $(m,3)$ DF TDBC Protocol is the union over all $\Delta_1+\Delta_2+\Delta_3=1, \Delta_1, \Delta_2, \Delta_3 \geq 0$ of

\begin{align}
R_\na &< \min_{\nr_i \in {\cal A}} \Delta_1 C(P|h_{\na,\nr_i}|^2) \\
R_\na &< \Delta_1 C(P|h_{\na,\nb}|^2) + \Delta_3 C\left(\sum_{\nr_i \in {\cal A}\cap {\cal B}} |h_{\nr_i,\nb}|^2 P_{{\cal A}\cap {\cal B}} + \sum_{\nr_j\in {\cal A}\setminus{\cal B}} |h_{\nr_j,\nb}|^2 P_{{\cal A}\setminus{\cal B}}  \right)\\
R_\nb &< \min_{\nr_i \in {\cal B}} \Delta_2 C(P|h_{\nb,\nr_i}|^2) \\
R_\nb &< \Delta_2 C(P|h_{\na,\nb}|^2) + \Delta_3 C\left(\sum_{\nr_i
\in {\cal A}\cap {\cal B}} |h_{\nr_i,\na}|^2 P_{{\cal A}\cap {\cal
B}} + \sum_{\nr_j\in {\cal B}\setminus{\cal A}} |h_{\nr_j,\na}|^2
P_{{\cal B}\setminus{\cal A}}  \right)
\end{align}
where $P_{{\cal A}\cap{\cal B}} = \frac{|{\cal A}\cap{\cal B}|}{|{\cal A}\cup{\cal B}|}P$, $P_{{\cal A}\setminus{\cal B}} = \frac{|{\cal A}\setminus{\cal B}|}{|{\cal A}\cup{\cal B}|}P$, and $P_{{\cal B}\setminus{\cal A}} =\frac{|{\cal B}\setminus{\cal A}|}{|{\cal A}\cup{\cal B}|}P$ over all ${\cal A, B}\subseteq {\cal R}$. \\

\item {\em $(m,m+2)$ DF MHMR Protocol}

The achievable rate region of the  $(m,m+2)$ DF MHMR Protocol is the union over all $\sum_{j=1}^{m+2}\Delta_j=1, \Delta_j \geq 0$ of  the rate pairs $(R_{\na}, R_{\nb})$ satisfying
\begin{align}
R_\na &< \min_{1\leq k\leq m+1} \left\{\sum_{i=1}^k \Delta_{m+3-i} C(P|h_{\nr_{i-1},\nr_k}|^2)\right\}\\
R_\nb &< \min_{1\leq k\leq m+1} \left\{\sum_{i=1}^k \Delta_{i}
C(P|h_{\nr_{m+2-i},\nr_{m+1-k}}|^2)\right\}
\end{align}

\item {\em $(m,t)$ DF MHMR Protocol}

For $3<t<m+2$, the achievable rate region will be the union over all $\sum_{j=1}^{t} \Delta_j = 1, \Delta_j\geq 0$ and all ${\cal R}_{j-1}\subset {\cal R}$ for $j\in[1,t]$ of the rate pairs $(R_{\na}, R_{\nb})$ satisfying
\begin{align}
R_\na &< \min_{1\leq k\leq t-1} \min_{\nr_k \in {\cal R}_k} \left\{\sum_{i=1}^k \Delta_{t+1-i} C\left(\sum_{\nr_{i-1} \in {\cal R}_{i-1}} |h_{\nr_{i-1},\nr_k}|^2 P\right)\right\}\\
R_\nb &< \min_{1\leq k\leq t-1} \min_{\nr_{t-1-k} \in {\cal
R}_{t-1-k}} \left\{\sum_{i=1}^k \Delta_{i} C\left(\sum_{\nr_{t-i}\in
{\cal R}_{t-i}}|h_{\nr_{t-i},\nr_{t-1-k}}|^2P\right)\right\}.
\end{align}
\end{itemize}

\subsection{Outer Bounds}
We derive outer bounds from Theorems \ref{theorem:MABC:out},
\ref{theorem:TDBC:out} and \ref{theorem:MHMR:1} in Gaussian
channel.

\begin{itemize}
\item {\em $(m,2)$ MABC Protocol}

The capacity region of the $(m,2)$ MABC Protocol is outer bounded  by the set of rate pairs $(R_{\na}, R_{\nb})$ satisfying
\begin{align}
R_\na &\leq \min_{S_R} \left\{\Delta_1 C\left(\sum_{\nr_i\in
\bar{S}_R} \frac{P}{2}|h_{\na,\nr_i}|^2\right) + \Delta_2
C\left(\sum_{\nr_i\in S_R} P|h_{\nr_i,\nb}|^2\right) \right\} \label{eq:mabc_g1}\\
R_\nb &\leq \min_{S_R} \left\{\Delta_1 C\left(\sum_{\nr_i\in
\bar{S}_R} \frac{P}{2}|h_{\nb,\nr_i}|^2\right) + \Delta_2
C\left(\sum_{\nr_i\in S_R} P|h_{\nr_i,\na}|^2\right) \right\},\label{eq:mabc_g2}
\end{align}
over all $\Delta_1+\Delta_2=1, \Delta_1, \Delta_2\geq 0$ and all
$S_R \subseteq {\cal R}$.

\item {\em $(m,3)$ TDBC Protocol}

The capacity region of the $(m,3)$ TDBC Protocol is outer bounded by the set of rate pairs $(R_{\na}, R_{\nb})$ satisfying
\begin{align}
R_\na &\leq \min_{S_R} \left\{\Delta_1 C\left(\sum_{\nr_i\in
\bar{S}_R} P|h_{\na,\nr_i}|^2 + P|h_{\na,\nb}|^2\right) + \Delta_3
C\left(\sum_{\nr_i\in S_R} P|h_{\nr_i,\nb}|^2\right) \right\}\\
R_\nb &\leq \min_{S_R} \left\{\Delta_2 C\left(\sum_{\nr_i\in
\bar{S}_R} P|h_{\nb,\nr_i}|^2 + P|h_{\nb,\na}|^2\right) + \Delta_3
C\left(\sum_{\nr_i\in S_R} P|h_{\nr_i,\na}|^2\right) \right\} \label{eq:tdbc_g2},
\end{align}
over all $\Delta_1+\Delta_2 + \Delta_3=1, \Delta_1, \Delta_2,
\Delta_3\geq 0$ and all $S_R \subseteq {\cal R}$.

\item {\em $(m,m+2)$ MHMR Protocol}

The capacity region of the $(m,m+2)$ MHMR Protocol is outer bounded by the set of rate pairs $(R_{\na}, R_{\nb})$ satisfying
\begin{align}
R_\na &\leq \min_{S_R} \left\{\sum_{\nr_i \in S_R\cup\{\na\}}
\Delta_{m+2-i} C\left(\sum_{\nr_j\in \bar{S}_R\cup\{\nb\}}
P|h_{\nr_i,\nr_j}|^2\right) \right\} \label{eq:mhmr_out:1}\\
R_\nb &\leq \min_{S_R} \left\{\sum_{\nr_i \in S_R\cup\{\nb\}}
\Delta_{m+2-i} C\left(\sum_{\nr_j\in \bar{S}_R\cup\{\na\}}
P|h_{\nr_i,\nr_j}|^2\right) \right\}, \label{eq:mhmr_out:2}
\end{align}
over all $\sum_{j=1}^{m+2}\Delta_j=1,  \Delta_j \geq 0$ and
all $S_R \subseteq {\cal R}$.
\end{itemize}


\section{Asymptotic analysis}

\label{sec:analysis}
We compare the sum rate $R_{\sum} \eqdef R_\na + R_\nb$ of the  proposed protocols and corresponding outer bounds at asymptotically low and high SNR (as $P \rightarrow 0$ and $P\rightarrow \infty$). To do so we employ the {\em Multiplicative Gap} as our comparison metric.
We define the Multiplicative Gap as the ratio between the achievable rate region and the outer bound of a given protocol, as follows:
\begin{definition}
\emph {Multiplicative Gap} $G_L$ and $G_H$ of a particular protocol are the ratio of the maximum achievable sum rate  ($R_{\sum}^{in}$) and the  sum rate outer bound ($R_{\sum}^{out}$) as $\text{SNR} \rightarrow 0$ and $\text{SNR} \rightarrow \infty$, respectively, i.e.,
\begin{align}
G_L &\eqdef \lim_{\text{SNR}\rightarrow 0} R_{\sum}^{in}/ R_{\sum}^{out} \\
G_H &\eqdef \lim_{\text{SNR}\rightarrow \infty} R_{\sum}^{in}/ R_{\sum}^{out}.
\end{align}
\end{definition}
From the definition, we note that $0\leq G_L, G_H \leq 1$: if $G_L$ (or $G_H$) approaches 1 we see that the inner and outer bounds tend to $\infty$ in the same fashion; if $G_L$ (or $G_H$) is close to 0 then the (additive) gap between inner and outer bounds increases with SNR.  While constant multiplicative gaps are more meaningful at low SNR, with constant additive gaps (between achievable rate regions and outer bounds), we consider high-SNR multiplicative gaps for two main reasons: 1) if the multiplicative gap $G_H$ is not equal to 1, such a scheme cannot hope to achieve a constant additive gap to capacity at high SNR -- this allows us to draw negative high-SNR conclusions about the various protocols, and 2) it is analytically tractable, holds for all channel gains and allows for an analytic comparison of the different protocols. We complement out high-SNR analysis with a low-SNR comparison of the DF protocols, and leave a more general comparison valid at all SNR for future work.

\subsection{Analytical protocol comparison at asymptotically low SNR}

In the very low SNR regime ($P \rightarrow 0$) $C(x) = \log_2 (1+x) \approx \frac{x}{\ln 2}$. We use this to approximate the achievable sum rates and outer bounds of Section \ref{sec:gaussian} at low SNR.  The achievable and outer bound sum-rates $R_{\sum}^{in}$ and $R_{\sum}^{out}$ at low SNR depend on the channel coefficients ($h_{i,j}$'s). However, we seek a comparison between the protocols which holds for a particular channel model without resorting to a series of channel gain-dependent scenarios. Thus, we will compare  upper and lower bounds of these sum rates for different protocols instead of comparing their exact values. We use the following notation:
\begin{align}
|h_{\max}| &\eqdef \max_{i,j\in {\cal R}^*} |h_{i,j}|\\
|h_{\min}| &\eqdef \min_{i,j\in {\cal R}^*} |h_{i,j}|
\end{align}
Also, we define $R_{\sum}(h)$ as the sum rate with every channel coefficient as $h$. Then, for the DF protocols
\begin{align}
R_{\sum}(h_{\min})\leq R_{\sum}\leq R_{\sum}(h_{\max})
\end{align}
as $\frac{\partial R_{\Sigma}}{\partial h_{i,j}} \geq 0, \;\; \forall h_{i,j}$. However, this property does not hold for the AF protocols, i.e. $ \exists h_{i,j}, \;\;\frac{\partial R_{\Sigma}}{\partial h_{i,j}} < 0$. Thus we only compare the relative performance with $G_L$ between the DF protocols and leave the comparison with the AF protocols for  future work.

We first derive upper and lower bounds of the $(m,2)$ DF MABC protocol. In \eqref{eq:dfmabc:1} $\sim$ \eqref{eq:dfmabc:5}, let us assume that all $h_{i,j}$'s are $h_{\max}$. Then,
\begin{align}
{\cal A} = {\cal B} = {\cal R}
\end{align}
and, using the approximation  $C(x) = \log_2 (1+x) \approx \frac{x}{\ln 2}$ valid at asymptotically low SNR,
\begin{align}
R_\na&< \min\left\{ \Delta_1 \frac{P}{2\ln 2}|h_{\max}|^2 , \Delta_2 \frac{mP}{\ln 2} |h_{\max}|^2\right\}\\
R_\nb&< \min\left\{ \Delta_1 \frac{P}{2\ln 2}|h_{\max}|^2 , \Delta_2 \frac{mP}{\ln 2} |h_{\max}|^2\right\}\\
R_\na + R_\nb &< \Delta_1 \frac{P}{2\ln 2} |h_{\max}|^2
\end{align}
Thus, optimizing $\Delta_1$ and $\Delta_2$ subject to $\Delta_1+\Delta_2=1$, we obtain $\Delta_1 = \frac{2m}{2m+1}$, $\Delta_2 = \frac{1}{2m+1}$ and
\begin{align}
R_{\sum}^{in} (h_{\max})  \approx  \frac{2m}{2m+1} \frac{P}{\ln 2} |h_{\max}|^2, \;\;\;\;\;\; R_{\sum}^{in} (h_{\min})  \approx \frac{2m}{2m+1} \frac{P}{\ln 2} |h_{\min}|^2.
\end{align}
Thus,
\begin{align}
\frac{2m}{2m+1} \frac{P}{\ln 2} |h_{\min}|^2 \leq R_{\sum}^{in} \leq \frac{2m}{2m+1} \frac{P}{\ln 2} |h_{\max}|^2
\end{align}
The upper and lower bounds of the $(m,2)$ MABC outer bound in \eqref{eq:mabc_g1} and \eqref{eq:mabc_g2} may be similarly derived to obtain: \begin{align}
\frac{2mP|h_{\min}|^2}{\ln 2} \leq R_{\sum}^{out} \leq \frac{2mP|h_{\max}|^2}{\ln 2}.
\end{align}
Therefore,
\begin{align}
\frac{1}{2m+1} \cdot \frac{|h_{\min}|^2}{|h_{\max}|^2}   \leq G_L \leq \frac{1}{2m+1}\cdot \frac{|h_{\max}|^2}{|h_{\min}|^2}.
\end{align}
 The other bounds can be similarly derived; we provide the results in Table \ref{table:low_SNR_sumrate} and the corresponding optimal (from a sum-rate perspective) $\Delta_i$'s in Table \ref{table:low_SNR_delta}. However, we use different upper and lower bounding techniques for the  $(m,m+2)$ MHMR outer bound. For a lower bound (or this outer bound) we simply apply equal time duration such that $\Delta_i = \frac{1}{m+2}$, $\forall i$ and find the sum rate. For the upper bound (of the outer bound) we optimize $\Delta_i$'s only for $R_\na$ \eqref{eq:mhmr_out:1} and simply multiply by 2 to obtain an upper bound.

\begin{table}
\caption{The upper and lower bounds of sum rates when $P \rightarrow 0$.}
\label{table:low_SNR_sumrate}
\centering
\begin{tabular}{l|c|c}
  \hline
  Protocol & $R_{\sum}(h_{\min})$ & $R_{\sum}(h_{\max})$ \\
  \hline
  $(m,2)$ DF MABC & $\frac{P|h_{\min}|^2}{\ln 2}\cdot \frac{2m}{2m+1} $ & $\frac{P|h_{\max}|^2}{\ln 2}\cdot \frac{2m}{2m+1} $\\
  $(m,2)$ MABC OUT & $\frac{2mP|h_{\min}|^2}{\ln 2}$ & $\frac{2mP|h_{\max}|^2}{\ln 2}$\\
  \hline
  $(m,3)$ DF TDBC & $\frac{P|h_{\min}|^2}{\ln 2}$ & $\frac{P|h_{\max}|^2}{\ln 2}$\\
  $(m,3)$ TDBC OUT & $\frac{2mP|h_{\min}|^2}{\ln 2}$ & $\frac{2mP|h_{\max}|^2}{\ln 2}$\\
  \hline
  $(m,m+2)$ DF MHMR & $\frac{P|h_{\min}|^2}{\ln 2}$ & $\frac{P|h_{\max}|^2}{\ln 2}$ \\
  $(m,m+2)$ MHMR OUT & $\frac{2P|h_{\min}|^2}{\ln 2}\cdot \frac{m+1}{m+2}$ & $\frac{2P|h_{\max}|^2}{\ln 2}$  \\
  \hline
\end{tabular}
\end{table}

\begin{table}
\caption{$\Delta_i$'s for upper and lower bounds when $P \rightarrow 0$ ($0\leq \alpha \leq 1$).}
\label{table:low_SNR_delta}
\centering
\begin{tabular}{l|c}
  \hline
  Protocol & $R_{\sum}(h_{\min})$ \\
  \hline
  $(m,2)$ DF MABC & $\Delta_1 =\frac{2m}{2m+1}$, $\Delta_2 = \frac{1}{2m+1}$ \\
  $(m,2)$ MABC OUT & $\Delta_1 =0$, $\Delta_2 = 1$  \\
  \hline
  $(m,3)$ DF TDBC & $\Delta_1 =\alpha$, $\Delta_2 = 1-\alpha$, $\Delta_3 = 0$ \\
  $(m,3)$ TDBC OUT & $\Delta_1 =0$, $\Delta_2 = 0$, $\Delta_3 = 1$  \\
  \hline
  $(m,m+2)$ DF MHMR & $\Delta_1 =\alpha$, $\Delta_{m+2} = 1-\alpha$, $\Delta_2 = \cdots =\Delta_{m+1} = 0$  \\
  $(m,m+2)$ MHMR OUT & $\Delta_1 =\cdots = \Delta_{m+2} =\frac{1}{m+2}$   \\
  \hline
  \hline
  Protocol & $R_{\sum}(h_{\max})$ \\
  \hline
  $(m,2)$ DF MABC &  $\Delta_1 =\frac{2m}{2m+1}$, $\Delta_2 = \frac{1}{2m+1}$\\
  $(m,2)$ MABC OUT & $\Delta_1 =0$, $\Delta_2 = 1$ \\
  \hline
  $(m,3)$ DF TDBC & $\Delta_1 =\alpha$, $\Delta_2 = 1-\alpha$, $\Delta_3 = 0$\\
  $(m,3)$ TDBC OUT &  $\Delta_1 =0$, $\Delta_2 = 0$, $\Delta_3 = 1$\\
  \hline
  $(m,m+2)$ DF MHMR &  $\Delta_1 =\alpha$, $\Delta_{m+2} = 1-\alpha$, $\Delta_2 = \cdots =\Delta_{m+1} = 0$  \\
  $(m,m+2)$ MHMR OUT &  $\Delta_1 =0$, $\Delta_i = \left(\frac{m+2-i}{i-1} -\frac{m+1-i}{i}\right)\frac{1}{m+1}$ for $2\leq i \leq m+1$, $\Delta_{m+2} = \frac{1}{m+1}$  \\
  \hline
\end{tabular}
\end{table}

The upper and lower bounds for multiplicative gaps in the very low SNR regime are
\begin{align}
 (m,2)~\text{DF MABC} &: \frac{1}{2m+1} \frac{|h_{\min}|^2}{|h_{\max}|^2} \leq G_L \leq \frac{1}{2m+1} \frac{|h_{\max}|^2}{|h_{\min}|^2}\\
 (m,3)~\text{DF TDBC} &: \frac{1}{2m} \frac{|h_{\min}|^2}{|h_{\max}|^2}  \leq G_L \leq \frac{1}{2m} \frac{|h_{\max}|^2}{|h_{\min}|^2}\\
 (m,m+2)~\text{DF MHMR} &: \frac{1}{2} \frac{|h_{\min}|^2}{|h_{\max}|^2}  \leq G_L \leq \frac{1}{2} \frac{|h_{\max}|^2}{|h_{\min}|^2}\frac{m+2}{m+1}.
\end{align}

In all protocols except for the $(m,m+2)$ DF MHMR protocol,  $G_L$ decreases with $m$. This may be interpreted as a decline in   performance of these protocols with network size.  However, for the $(m,m+2)$ DF MHMR protocol, $G_L$ is independent of the size of the network and is the largest of all the presented protocols, from which we may conclude that  the $(m,m+2)$ DF MHMR protocol is the best protocol in this channel regime.

\subsection{Analytical protocol comparison at asymptotically high SNR}

In the very high SNR regime ($P \rightarrow \infty$) $C(x) = \log_2 (1+x) \approx \log_2 x$. We use this approximation to derive the achievable sum rates and outer bounds of Section \ref{sec:gaussian} at high SNR, for both the DF and AF versions of the protocols. We include the derivation of upper and lower bounds of $(m,2)$ AF MABC protocol case here as an example. Other cases are similarly derived.

In \eqref{eq:afmabc:1} and \eqref{eq:afmabc:2}, we have,
\begin{align}
R_\na &< \frac12 C(\alpha_1 P) \approx \frac12 (\log P + \log \alpha_1)\\
R_\nb &< \frac12 C(\alpha_2 P) \approx \frac12 (\log P + \log \alpha_2),
\end{align}
where $\alpha_1$ and $\alpha_2$ are constants determined by $m$, $\{|h_{\nb,\nr_i}|\}$ and $\{|h_{\na,\nr_i}|\}$. Thus we have,
\begin{align}
R_{\sum} < \log P  + \log \alpha_1 \alpha_2.
\end{align}
Since $\alpha_1$ and $\alpha_2$ are given constants, $R_{\sum} \rightarrow \log P + \texttt{constant}$ as $P\rightarrow \infty$.

The other multiplicative gains are derived similarly; they are summarized in Table \ref{table:high_SNR_sumrate}. The multiplicative gaps in the very high SNR regime are thus
\begin{align}
 (m,2)~\text{AF MABC} : G_H &= \frac12\\
 (m,2)~\text{DF MABC} : G_H &= \frac13\\
 (m,3)~\text{AF TDBC} : G_H &= \frac13\\
 (m,3)~\text{DF TDBC} : G_H &= \frac12\\
 (m,m+2)~\text{AF MHMR} : G_H &= \frac{2}{m+2}\\
 (m,m+2)~\text{DF MHMR} : G_H &= 1.
\end{align}
From the summary in Table \ref{table:high_SNR_sumrate} the $(m,2)$ AF MABC, $(m+3)$ DF TDBC and $(m,m+2)$ DF MHMR protocol perform ``better'' in the very high SNR regime, while the $(m,m+2)$ AF MHMR protocol is the worst. The corresponding optimized $\Delta_i$'s are summarized in Table \ref{table:high_SNR_delta}. These results match the intuition and conclusions that we will draw in the next section on numerical results under very specific channel conditions. Notably, the $(m,m+2)$ DF MHMR protocol achieves a high-SNR multiplicative gap of 1, thus motivating its use at high SNR.

\begin{table}
\caption{sum rates when $P \rightarrow \infty$. $\{C_i\}$ are the corresponding constant terms.}
\label{table:high_SNR_sumrate}
\centering
\begin{tabular}{l|c}
  \hline
  Protocol & sum rate  \\
  \hline
  $(m,2)$ AF MABC & $\log P + C_1$  \\
  $(m,2)$ DF MABC & $\frac23 \log P + C_2 $ \\
  $(m,2)$ MABC OUT & $2 \log P + C_3$\\
  \hline
  $(m,3)$ AF TDBC & $\frac23 \log P + C_4$ \\
  $(m,3)$ DF TDBC & $\log P + C_5 $  \\
  $(m,3)$ TDBC OUT & $2 \log P + C_6$\\
  \hline
  $(m,m+2)$ AF MHMR & $\frac2{m+2} \log P + C_7$ \\
  $(m,m+2)$ DF MHMR & $\log P + C_8 $ \\
  $(m,m+2)$ MHMR OUT & $\log P + C_9$\\
  \hline
\end{tabular}
\end{table}

\begin{table}
\caption{$\Delta_i$'s when $P \rightarrow \infty$ ($0\leq \alpha \leq 1$).}
\label{table:high_SNR_delta}
\centering
\begin{tabular}{l|c}
  \hline
  Protocol & $\Delta_i$'s \\
  \hline
  $(m,2)$ DF MABC & $\Delta_1 =\frac{2}{3}$, $\Delta_2 = \frac{1}{3}$ \\
  $(m,2)$ MABC OUT & $\Delta_1 =\alpha$, $\Delta_2 = 1-\alpha$  \\
  \hline
  $(m,3)$ DF TDBC & $\Delta_1 =\alpha$, $\Delta_2 = 1-\alpha$, $\Delta_3 = 0$ \\
  $(m,3)$ TDBC OUT & $\Delta_1 =0$, $\Delta_2 = 0$, $\Delta_3 = 1$  \\
  \hline
  $(m,m+2)$ DF MHMR & $\Delta_1 =\alpha$, $\Delta_{m+2} = 1-\alpha$, $\Delta_2 = \cdots =\Delta_{m+1} = 0$  \\
  $(m,m+2)$ MHMR OUT & $\Delta_1 =\alpha$, $\Delta_{m+2} = 1-\alpha$, $\Delta_2 = \cdots =\Delta_{m+1} = 0$   \\
  \hline
\end{tabular}
\end{table}


\section{Numerical analysis}

\label{sec:regions}

\subsection{Rate region comparisons with one to two relays}

\begin{figure*}
  \hfill
  \begin{minipage}[t]{.45\textwidth}
    \begin{center}
       \epsfig{figure=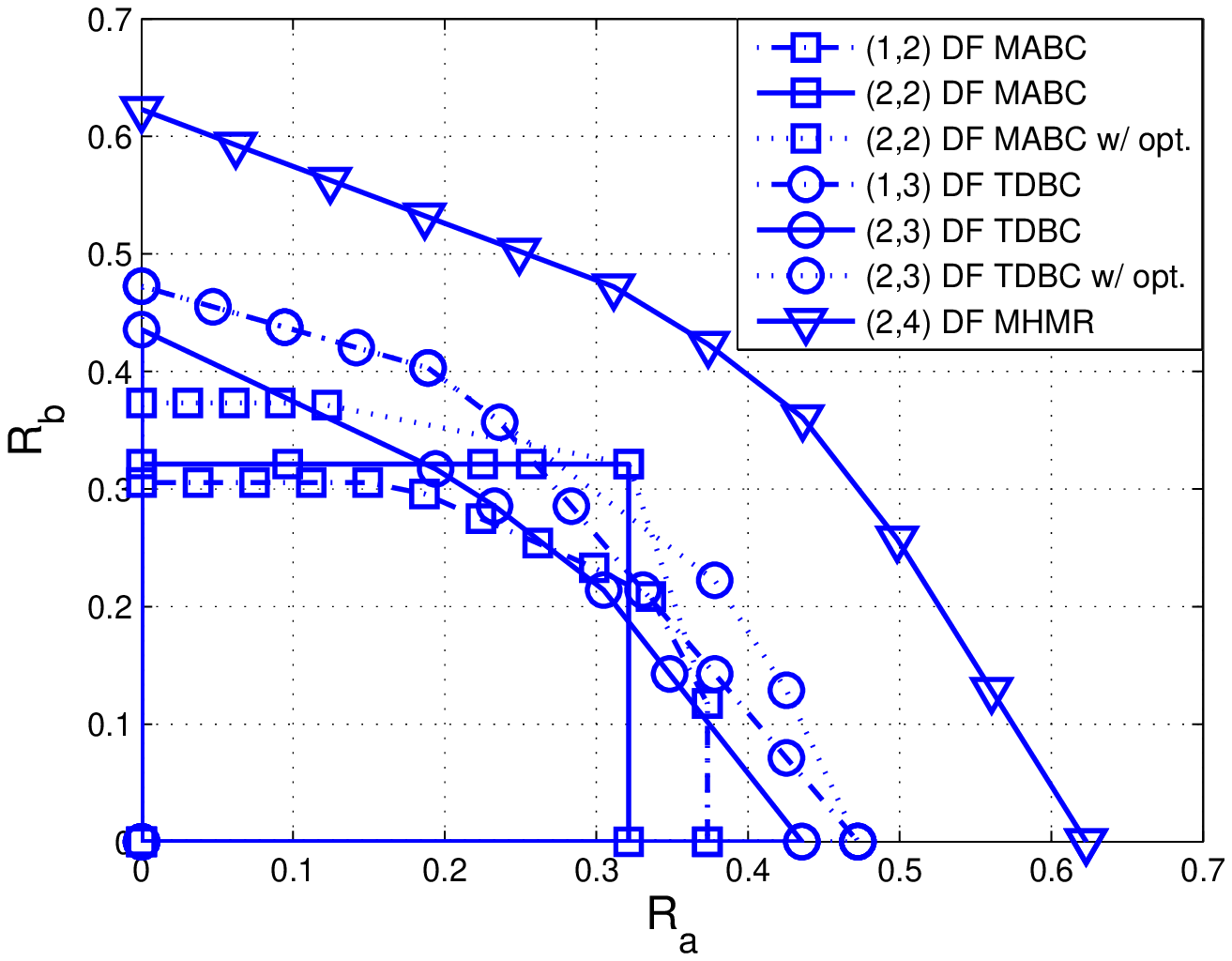, width=7.5cm}
      \caption{\baselineskip=10pt Comparison of achievable regions between single relay and multiple relays with $P=0$ dB.}
            \label{fig:comp_single_low}
    \end{center}
  \end{minipage}
  \hfill
  \begin{minipage}[t]{.45\textwidth}
    \begin{center}
        \epsfig{figure=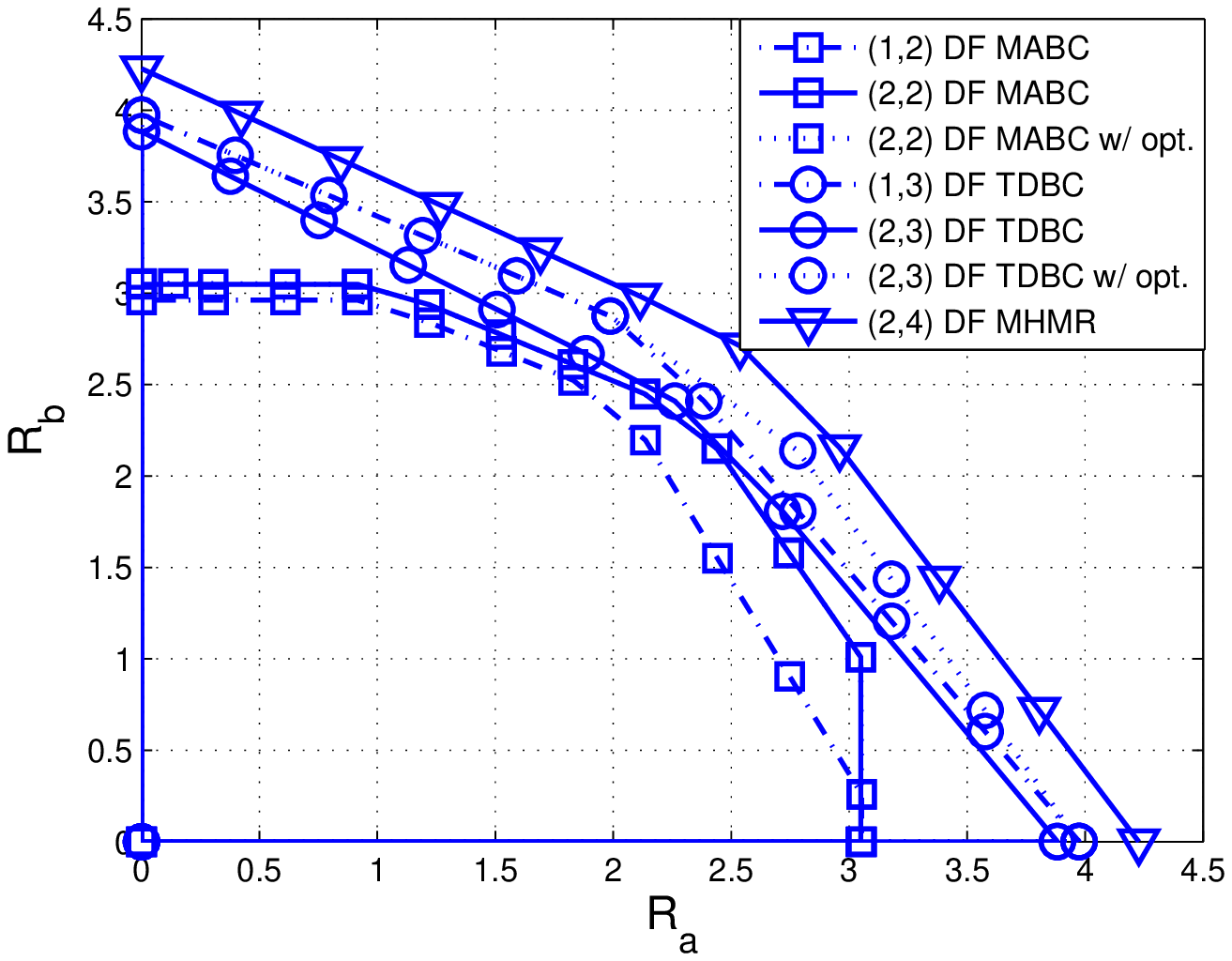, width=7.5cm}
      \caption{\baselineskip=10pt Comparison of achievable regions between single relay and multiple relays with $P=20$ dB.}
      \label{fig:comp_single_high}
    \end{center}
  \end{minipage}
  \hfill
\end{figure*}

\begin{figure*}
  \hfill
  \begin{minipage}[t]{.45\textwidth}
    \begin{center}
       \epsfig{figure=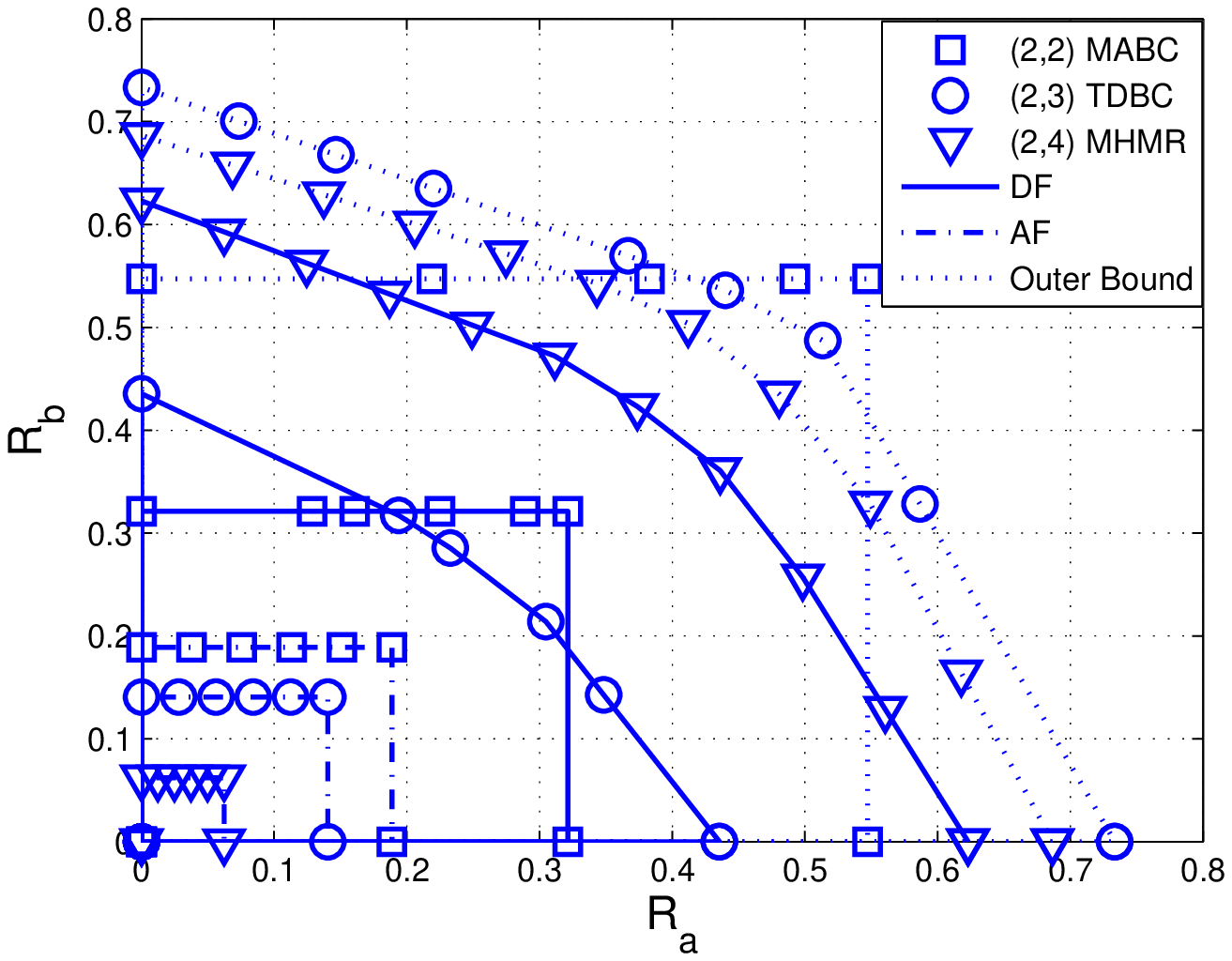, width=7.5cm}
      \caption{\baselineskip=10pt Comparison of achievable regions of AF and DF and outer bounds with $P=0$ dB.}
            \label{fig:comp_af_low}
    \end{center}
  \end{minipage}
  \hfill
  \begin{minipage}[t]{.45\textwidth}
    \begin{center}
        \epsfig{figure=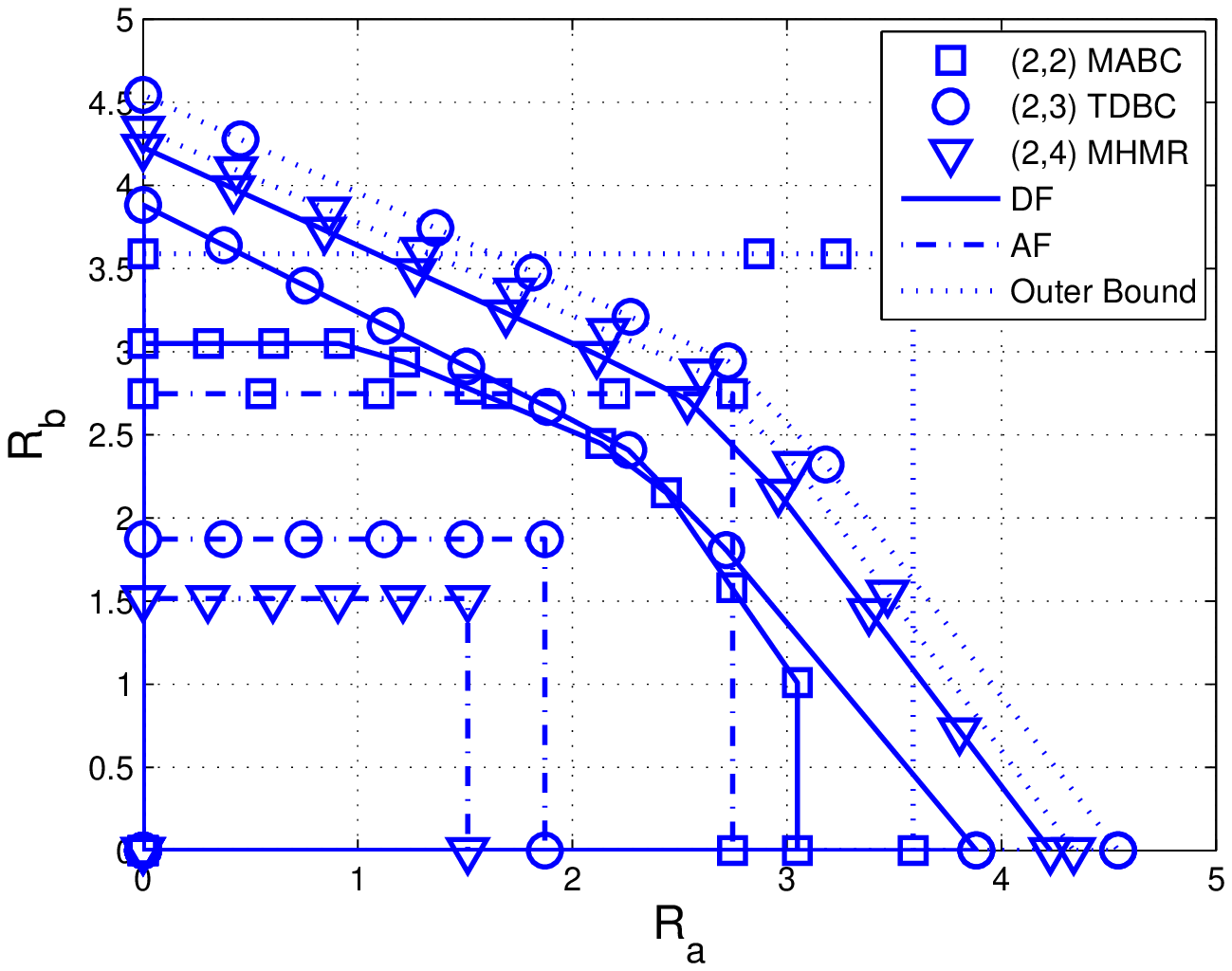, width=7.5cm}
      \caption{\baselineskip=10pt Comparison of achievable regions of AF and DF and outer bounds with $P=20$ dB.}
      \label{fig:comp_af_high}
    \end{center}
  \end{minipage}
  \hfill
\end{figure*}

In this section we numerically evaluate the rate regions obtained in  the previous section for a variety of parameters, which include the number of relays, the type of relaying (DF or AF), as well as the number of hops $t$ and whether these hops are regular.  This complements the low and high SNR analysis obtained in the previous section. Specifically, we look at:
\begin{itemize}
\item {\it One relay versus two relays under with DF relaying:} We compare the achievable regions of two single relay protocols (MABC and TDBC) and three two-relay MHMR protocols with DF schemes at low (\Fig \ref{fig:comp_single_low}) and high (Fig. \ref{fig:comp_single_high}) SNRs.
\item {\it DF versus AF relaying:} We compare the regions of DF and AF relaying in the MHMR protocols  at low (\Fig \ref{fig:comp_af_low}) and high (Fig. \ref{fig:comp_af_high}) SNRs.
\end{itemize}

We use the following channel gain matrix \footnote{If other channel gains are chosen, the numerical results may change; however some general statements may be made regardless, as supported by the analytical results at low and high SNR.}:
\begin{align}
{\bf H} = \left[
            \begin{array}{cccc}
              0 & 1.2 & 0.8 & 0.2 \\
              1.2 & 0 & 2 & 0.8 \\
              0.8 & 2 & 0 & 1.2 \\
              0.2 & 0.8 & 1.2 & 0 \\
            \end{array}
          \right]
\end{align}

In the DF relaying protocols, the (2,4) MHMR protocol outperforms
the other protocols at both low and high SNR. This improved
performance may be attributed to this protocol's  effective use of
side information.
During each phase, every node which is not transmitting can receive the current transmission which it may employ as side information to aid decoding during later stages. It may also subtract off the part of the transmission corresponding to the  message(s) it already knows.
There is naturally a tradeoff between the number of phases and the
amount of information broadcasted in each phase. However, as seen by
our simulations in this particular channel,  the effect of reducing
the number of phases to 2 or 3 does not outweigh the effect of
broadcasting information.

It is interesting to note that the (1,2) DF MABC and (1,3) DF TDBC
protocols  may outperform the (2,2) DF MABC and (2,3) DF TDBC
protocols in some scenarios. This reveals, as suspected, that using one relay and
allocating all transmit power to that single node is sometimes
better than using multiple relays with equal power allocated to each
of them. However, if we allow power optimization between
different subsets of relays then multiple relaying protocols
outperform the single relay protocols (dotted lines in \Fig
\ref{fig:comp_single_low} and \ref{fig:comp_single_high}). This
reveals that we achieve larger gain if power allocation between the
relays participating in the transmission of messages is permitted.

The inner and outer bounds differ for a number of reasons, with the prevailing one being that our inner bounds use a DF scheme. For the MABC scheme using DF relaying,  in equations \eqref{eq:mabc_g1}--\eqref{eq:tdbc_g2}, every relay contributes to  enlarging the outer bound regions, while only the subset of relays ${\cal A} \cup {\cal B}$ are used in determining the achievable regions.  At low SNR, when ${\cal A}\cup {\cal B}$ is relatively small, the gaps, shown in \Fig \ref{fig:comp_af_low} are larger than the gaps at high SNRs shown in \Fig \ref{fig:comp_af_high}, where the number of relays in ${\cal A}\cup {\cal B}$ are relatively larger. In addition to simply having more relays contribute to the outer bound regions, their effect is summed up outside of the logarithm for the outer bound, and inside of it for the inner bounds. For example, in the MABC protocol, $R_\na \leq \Delta_1 C(\cdot) + \Delta_2 C(\cdot)$ for the outer bound,  as opposed to $R_\na \leq C(\sum \cdot)$ for the inner bound.
  Lastly, the achievable rate regions for DF relaying are significantly reduced by the necessity of  having all relays decode the message(s) $w_\na$ or $w_\nb$ individually, resulting in the min function which significantly diminishes the region. This requirement to decode all messages is not present in the outer bounds. The inner bounds for the AF relaying schemes are relatively small as (a) noise is carried forward, (b) no power optimization is performed  and (c) no phase-length  optimization is performed. The inner bounds may be improved through the use of compress and forward relaying \cite{SKim:2008a} or de-noising, which may be able to capture the optimal tradeoff between eliminating the noise while not requiring the messages to be decoded. The exploration of different relaying schemes as well as the analytical  impact of different channel gain matrices is left for future work.

In the proposed protocols, the (2,4) DF MHMR protocol achieves the largest rate region in most scenarios. In the high SNR regime, the (2,2) AF MABC protocol may achieve rates slightly better  than the (2,4) DF MHMR protocol, as noise amplification is less of an issue. Furthermore, the (2,2) AF MABC protocol outperforms the  (2,3) AF TDBC protocol since it employs less phases and the interference is perfectly canceled at each terminal node.
 However as a general rule, multiple hops with DF relaying is the optimum protocol in  this bi-directional half-duplex channel.

\subsection{Rate region comparison with 8  relays on a line.}
In this subsection 8 relays are placed on the line between $\na$ and
$\nb$. The distance from $\na$ to $\nr_i$ is $d_{\na\nr} =
\frac{i}{9} d_{\na\nb}$ ($1\leq i\leq 8$). Thus, $d_{\nb\nr} =
(1-\frac{i}{9}) d_{\na\nb}$. We let $h_{\na\nb} = 0.2$ and
$|h_{ij}|^2 = k  /d_{ij}^{3.8}$ for $k$ constant and a path-loss
exponent of $3.8$.

\begin{figure*}
  \hfill
  \begin{minipage}[t]{.45\textwidth}
    \begin{center}
       \epsfig{figure=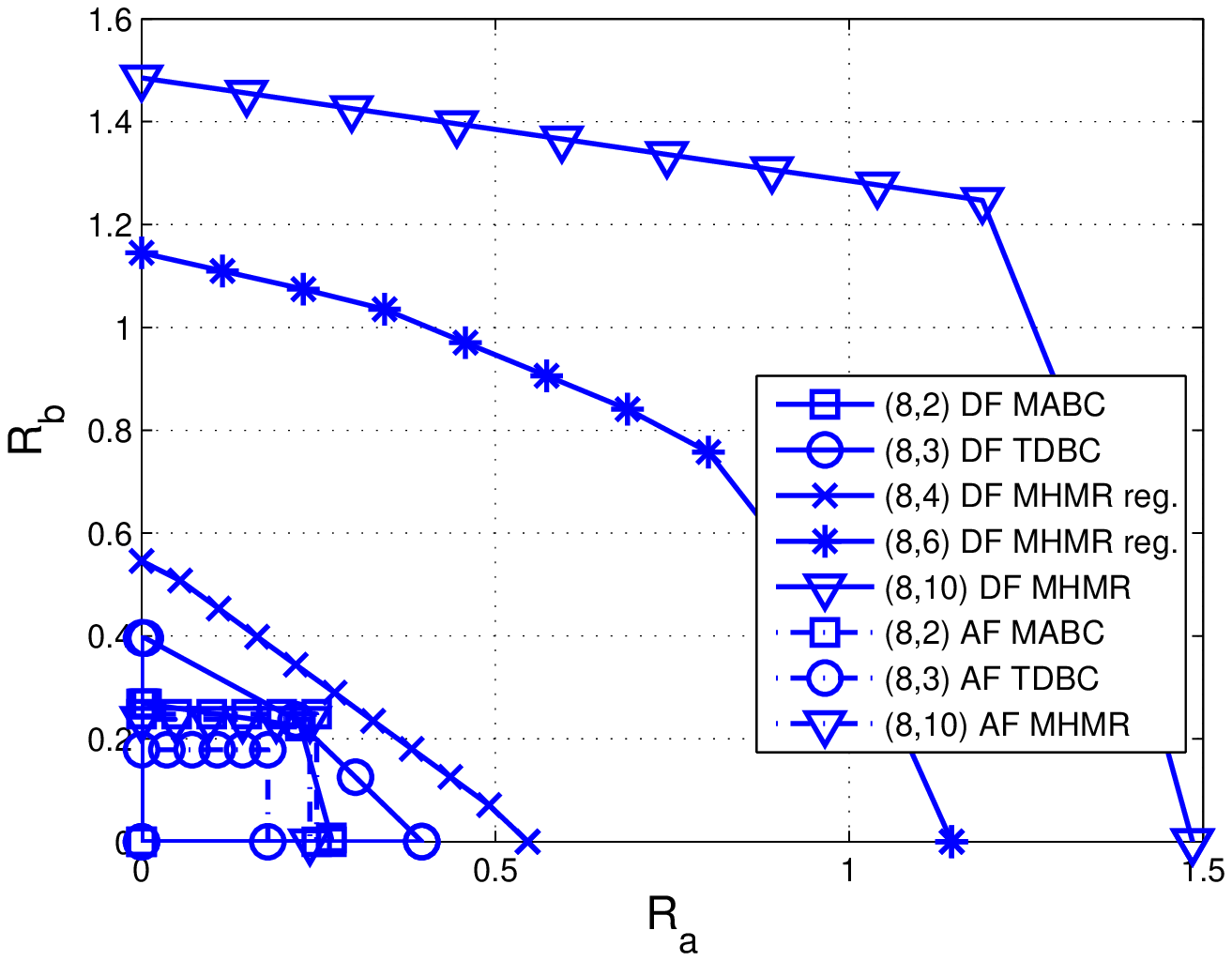, width=7.5cm}
      \caption{\baselineskip=10pt Comparison of achievable regions with 8  relays on a line with $P=0$ dB.}
            \label{fig:line_comp_low}
    \end{center}
  \end{minipage}
  \hfill
  \begin{minipage}[t]{.45\textwidth}
    \begin{center}
        \epsfig{figure=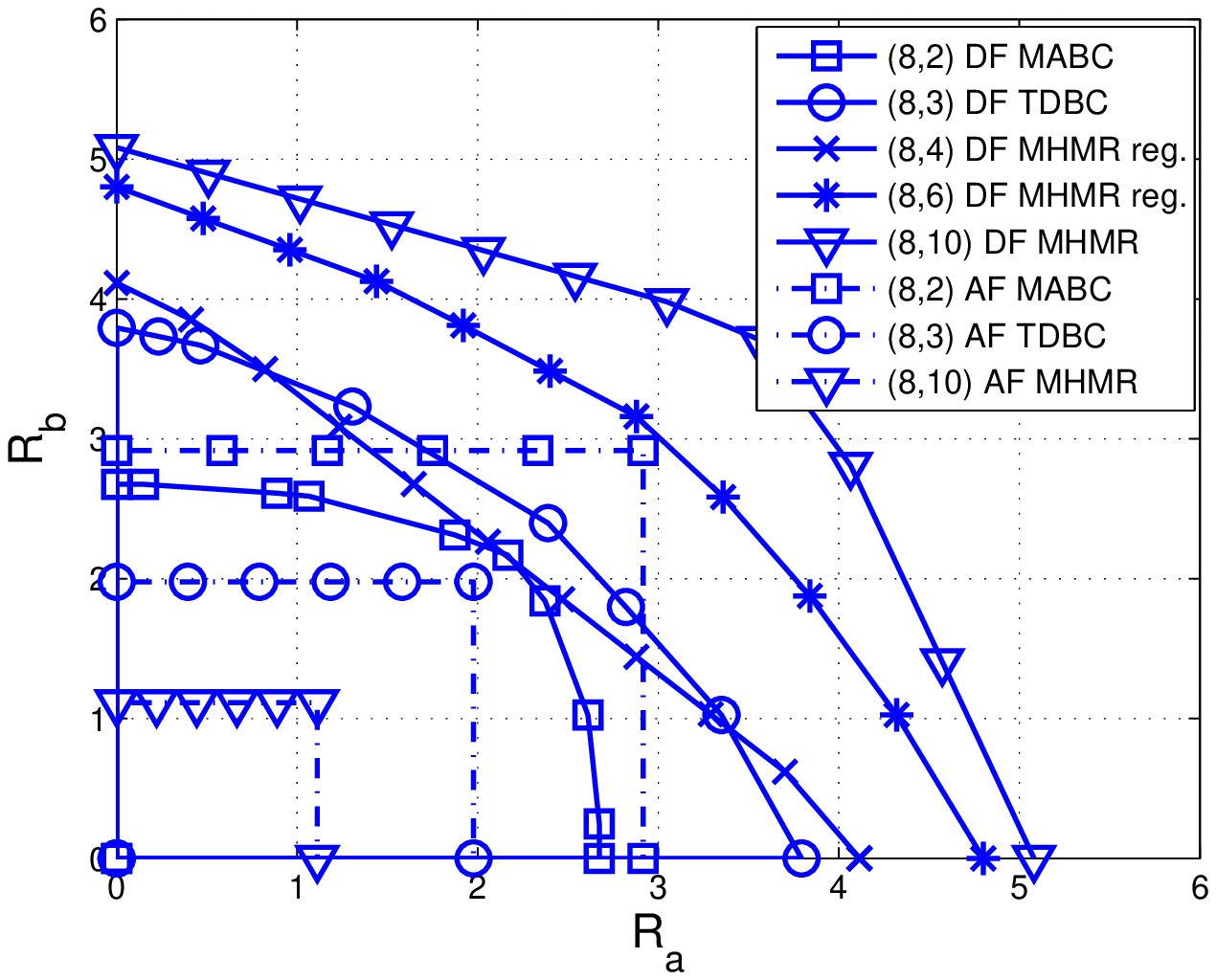, width=7.5cm}
      \caption{\baselineskip=10pt Comparison of achievable regions with 8  relays on a line with $P=20$ dB.}
      \label{fig:line_comp_high}
    \end{center}
  \end{minipage}
  \hfill
\end{figure*}

In \Fig \ref{fig:line_comp_low} and \ref{fig:line_comp_high},
the (8,10) DF MHMR protocol dominates the other protocols both in
the low SNR and high SNR regime. As we explained in the previous
subsection, this may be attributed to the  broadcasted side
information. While increasing the number of phases means that  less
information may be transmitted  during each time phase, the
accumulated side information and improved channel gains (shorter
distances) for each hop outweighs these detrimental effects,
yielding higher overall rates.

In contrast to the DF scheme, the achievable rate region for the AF
schemes decreases as the number of hops increases, as the noise is
increasingly amplified and carried forward. Similarly,  in the low
SNR regime (\Fig \ref{fig:line_comp_low}) when the noise is very
large to begin with, as expected, the AF schemes performs very
poorly.

\subsection{Sum-rate with an increasing number of relays}

We consider the same geometric location of the relays as  in the previous subsection but increase their number. We compare the optimized  sum rates $(R_\na + R_\nb)$ of the different relaying schemes.

\begin{figure*}
  \hfill
  \begin{minipage}[t]{.45\textwidth}
    \begin{center}
       \epsfig{figure=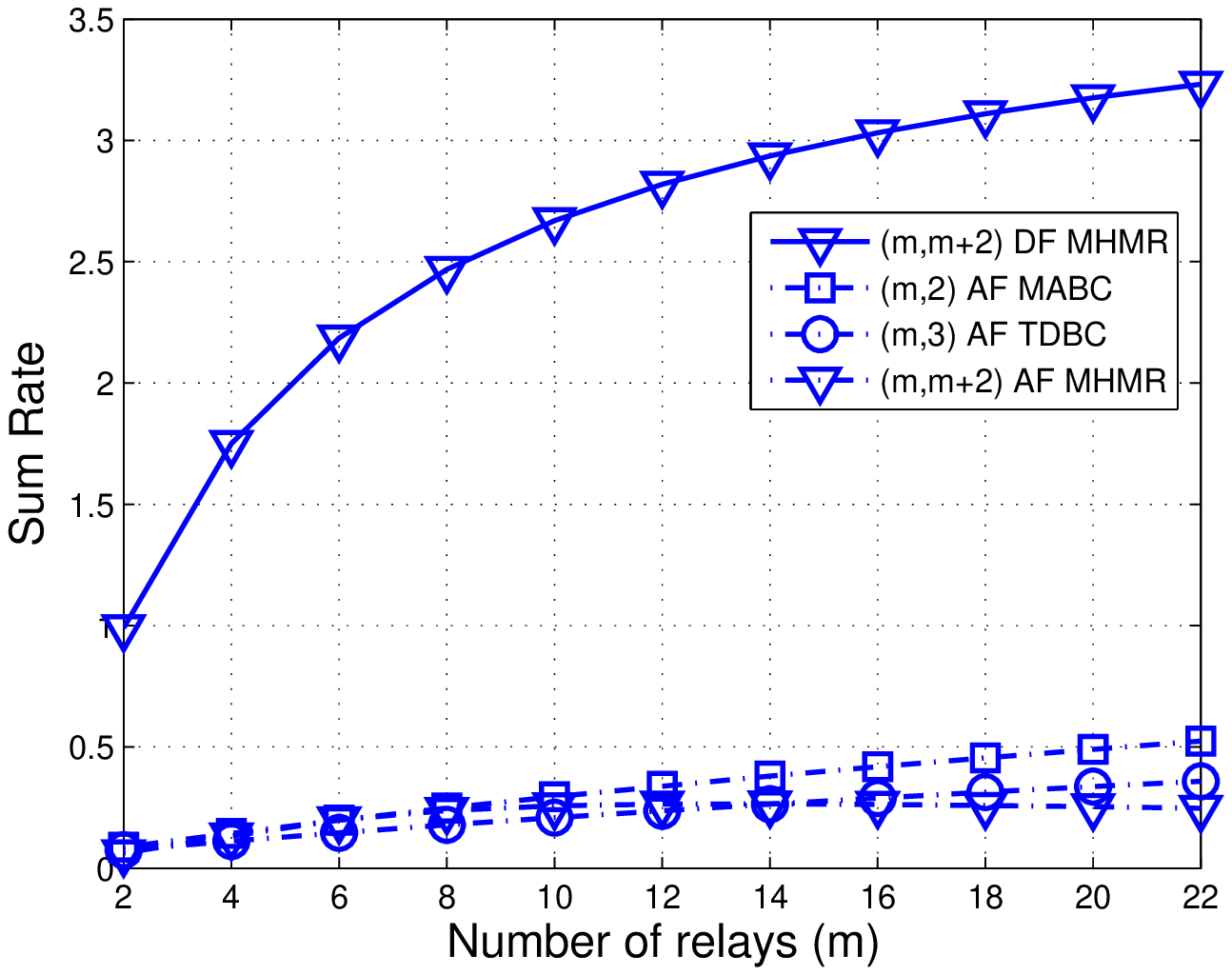, width=7.5cm}
      \caption{\baselineskip=10pt Comparison of achievable sum rates with $P=0$ dB.}
            \label{fig:relay_num_low}
    \end{center}
  \end{minipage}
  \hfill
  \begin{minipage}[t]{.45\textwidth}
    \begin{center}
        \epsfig{figure=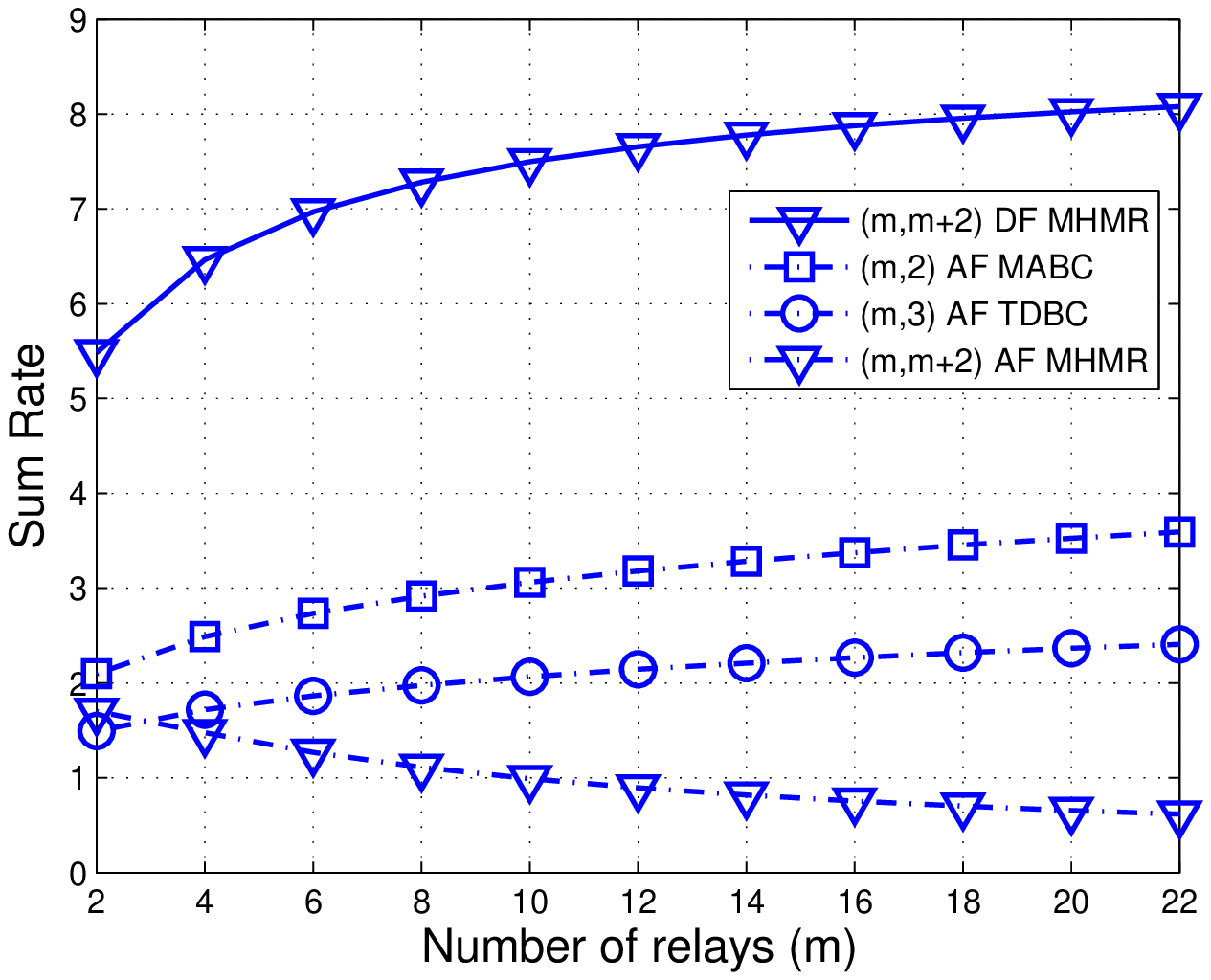, width=7.5cm}
      \caption{\baselineskip=10pt Comparison of achievable sum rates with $P=20$ dB.}
      \label{fig:relay_num_high}
    \end{center}
  \end{minipage}
  \hfill
\end{figure*}

In \Fig \ref{fig:relay_num_low} and \ref{fig:relay_num_high}, the
$(m,m+2)$ DF MHMR protocol outperforms the other protocols. Also,
with more relays, the $(m,m+2)$ DF MHMR protocol improves its performance, while the $(m,m+2)$ AF MHMR protocol's performance deteriorates. The
tendency can be seen more significantly in the high SNR regime (\Fig
\ref{fig:relay_num_high}).

\subsection{Sum-rate with two relays on a line}
We consider the case that two relays $\nr_1$ and $\nr_2$ are located on a line between $\na$ and $\nb$. We assume that $0<d_{\na\nr_1} < d_{\na\nr_2} < d_{\na\nb} = 1$ with path-loss exponent 3.8 and transmit power 0 dB. We plot the sum-rate achieved by the different AF and DF protocols for all values of the $0<d_{\na\nr_1} < d_{\na\nr_2} < d_{\na\nb} = 1$, resulting in 3-D plots; color is useful in understanding these plots.

\begin{figure*}
  \hfill
  \begin{minipage}[t]{.45\textwidth}
    \begin{center}
       \epsfig{figure=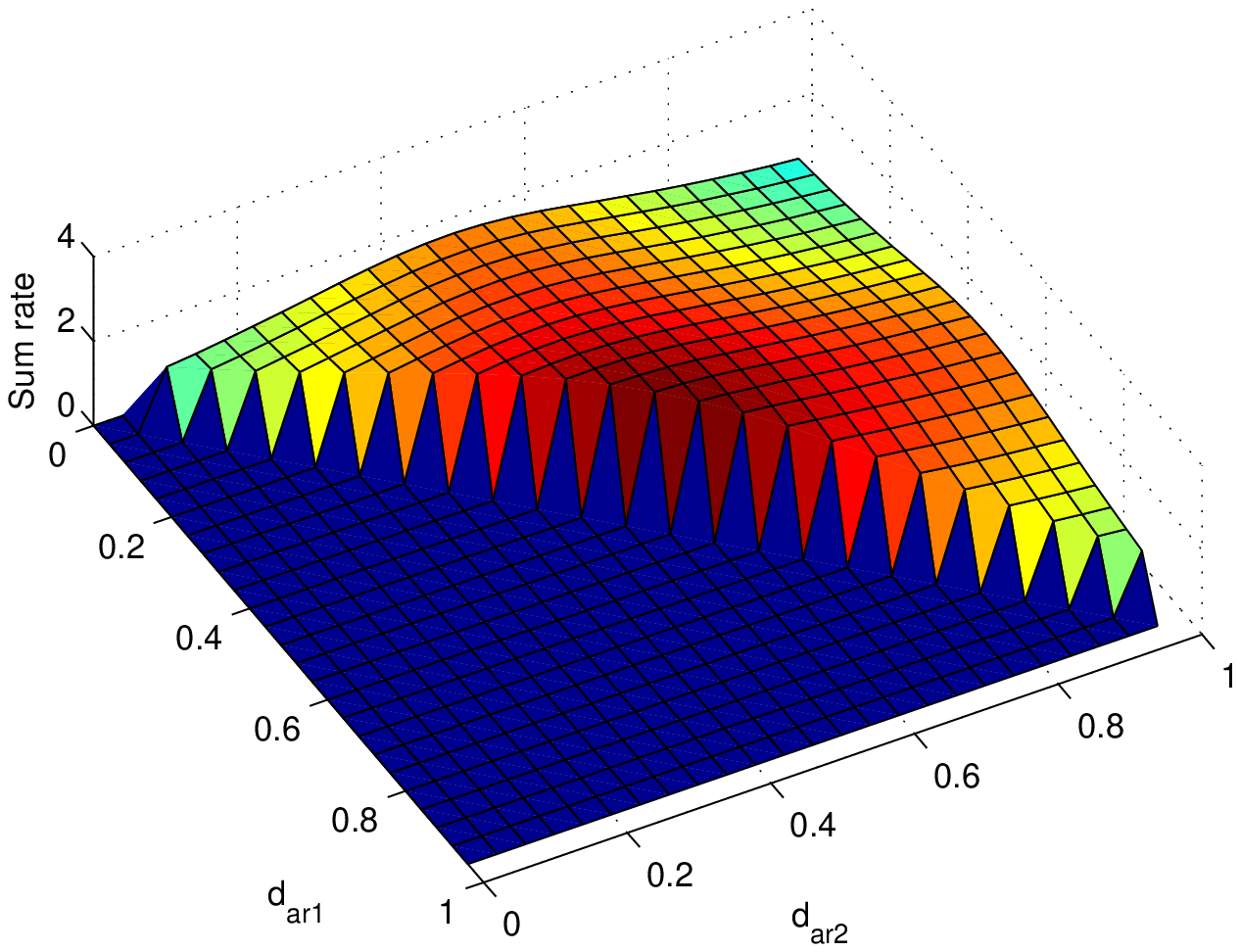, width=8cm}
      \caption{\baselineskip=10pt achievable sum rates of the $(2,2)$ AF MABC protocol with $P=0$ dB.}
            \label{fig:two_line_af_mabc}
    \end{center}
  \end{minipage}
  \hfill
  \begin{minipage}[t]{.45\textwidth}
    \begin{center}
        \epsfig{figure=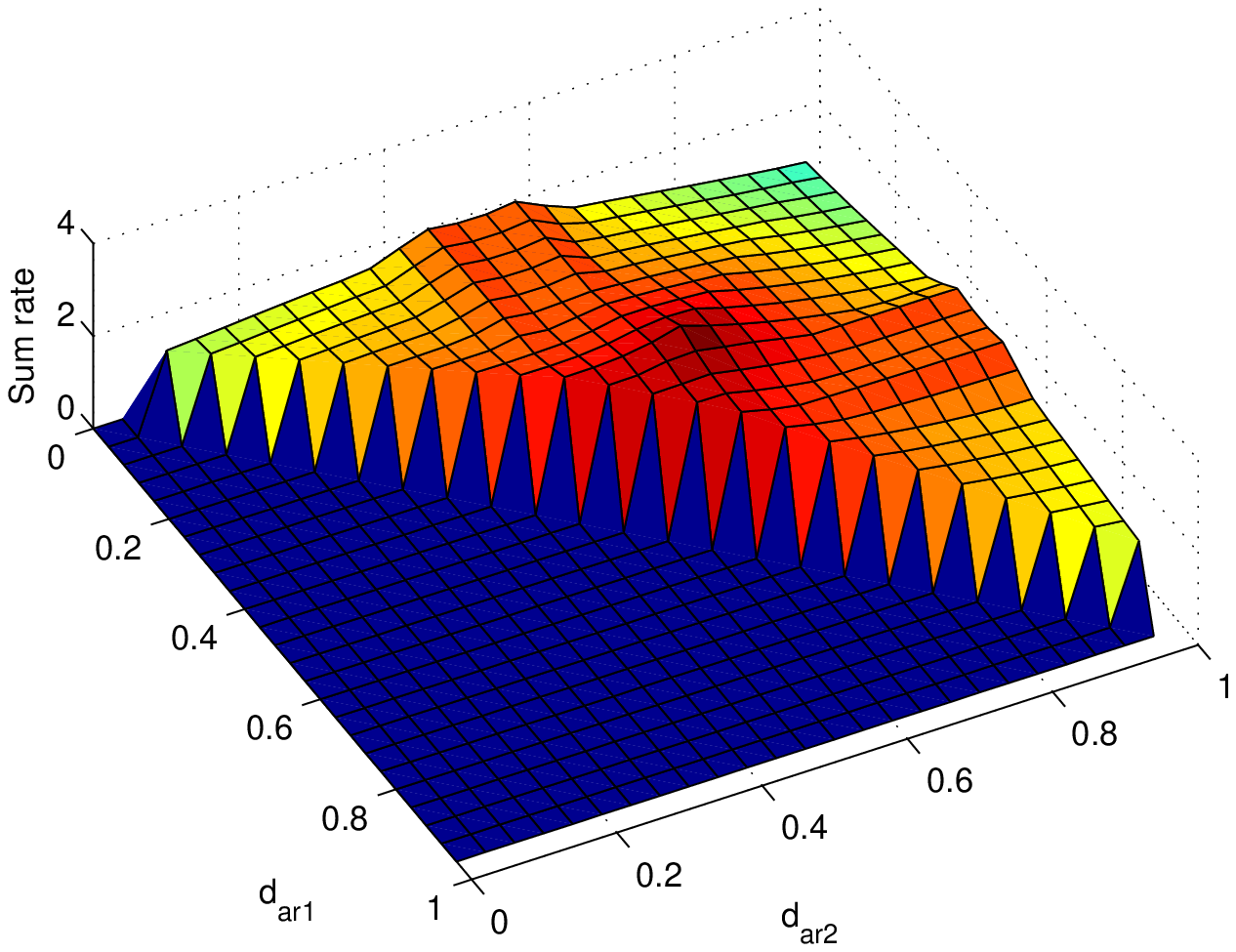, width=8cm}
      \caption{\baselineskip=10pt achievable sum rates of the $(2,2)$ DF MABC protocol with $P=0$ dB.}
      \label{fig:two_line_df_mabc}
    \end{center}
  \end{minipage}
  \hfill
\end{figure*}

\begin{figure*}
  \hfill
  \begin{minipage}[t]{.45\textwidth}
    \begin{center}
       \epsfig{figure=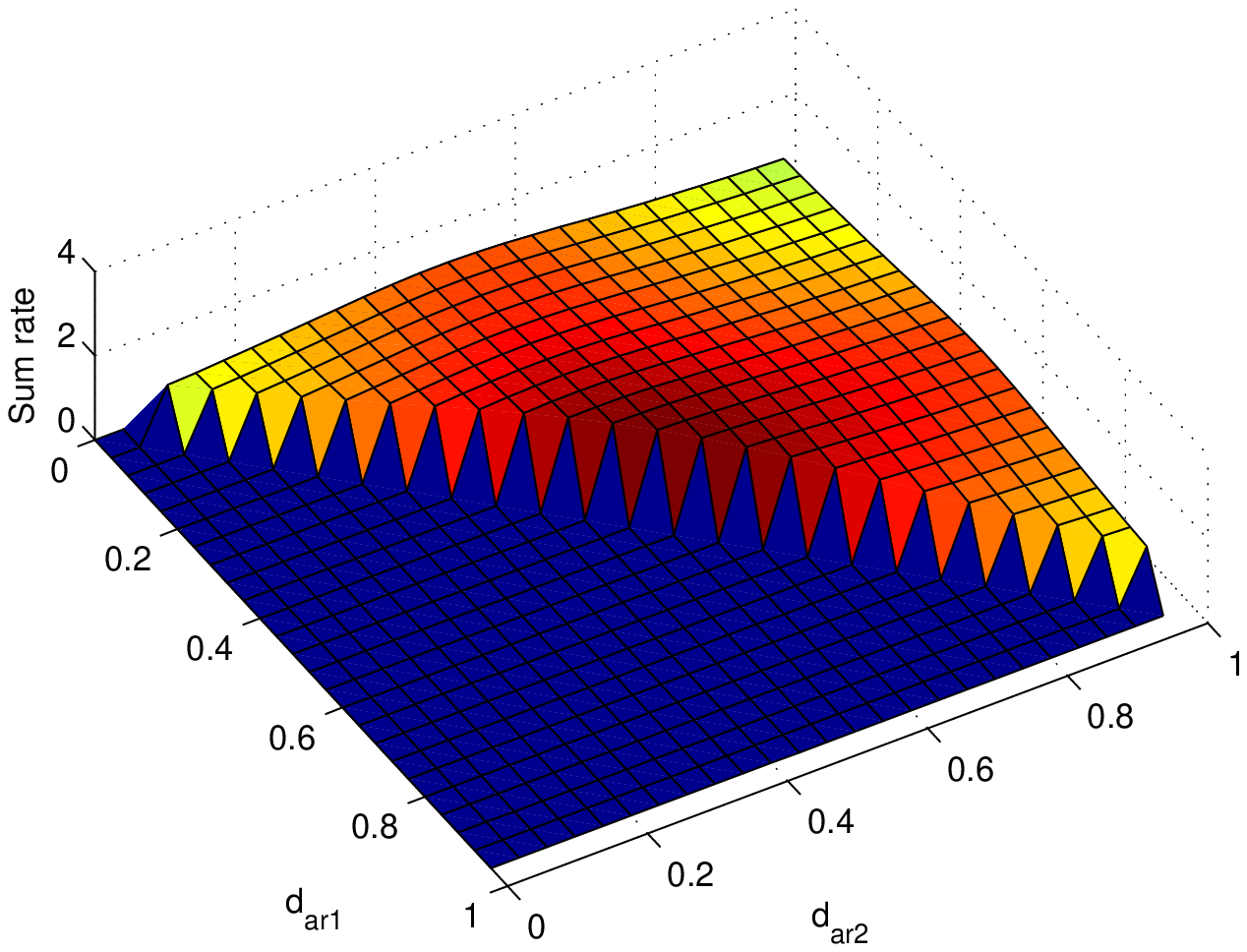, width=8cm}
      \caption{\baselineskip=10pt achievable sum rates of the $(2,4)$ AF TDBC protocol with $P=0$ dB.}
            \label{fig:two_line_af_tdbc}
    \end{center}
  \end{minipage}
  \hfill
  \begin{minipage}[t]{.45\textwidth}
    \begin{center}
        \epsfig{figure=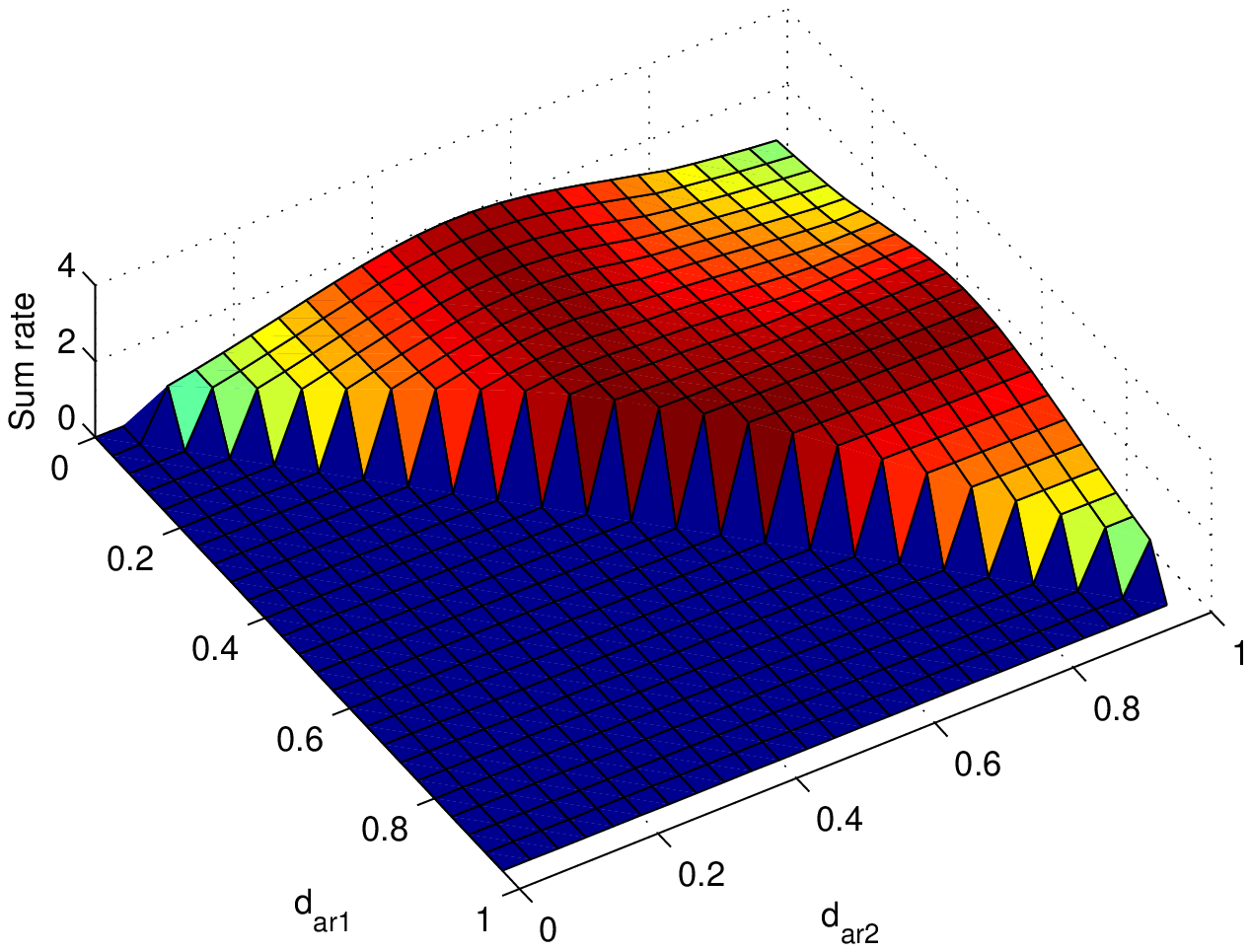, width=8cm}
      \caption{\baselineskip=10pt achievable sum rates of the $(2,3)$ DF TDBC protocol with $P=0$ dB.}
      \label{fig:two_line_df_tdbc}
    \end{center}
  \end{minipage}
  \hfill
\end{figure*}

\begin{figure*}
  \hfill
  \begin{minipage}[t]{.45\textwidth}
    \begin{center}
       \epsfig{figure=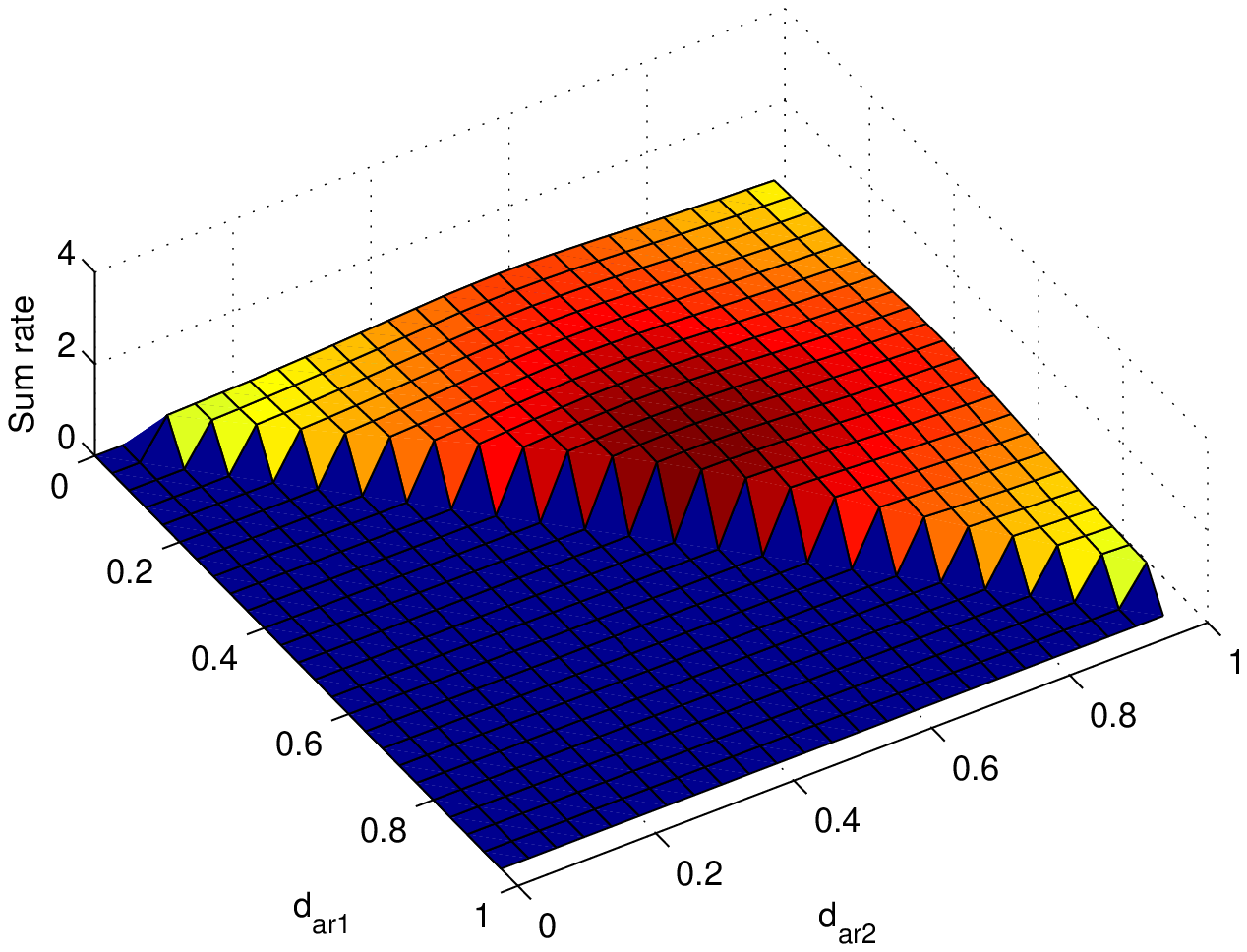, width=8cm}
      \caption{\baselineskip=10pt achievable sum rates of the $(2,4)$ AF MHMR protocol with $P=0$ dB.}
            \label{fig:two_line_af_mhmr}
    \end{center}
  \end{minipage}
  \hfill
  \begin{minipage}[t]{.45\textwidth}
    \begin{center}
        \epsfig{figure=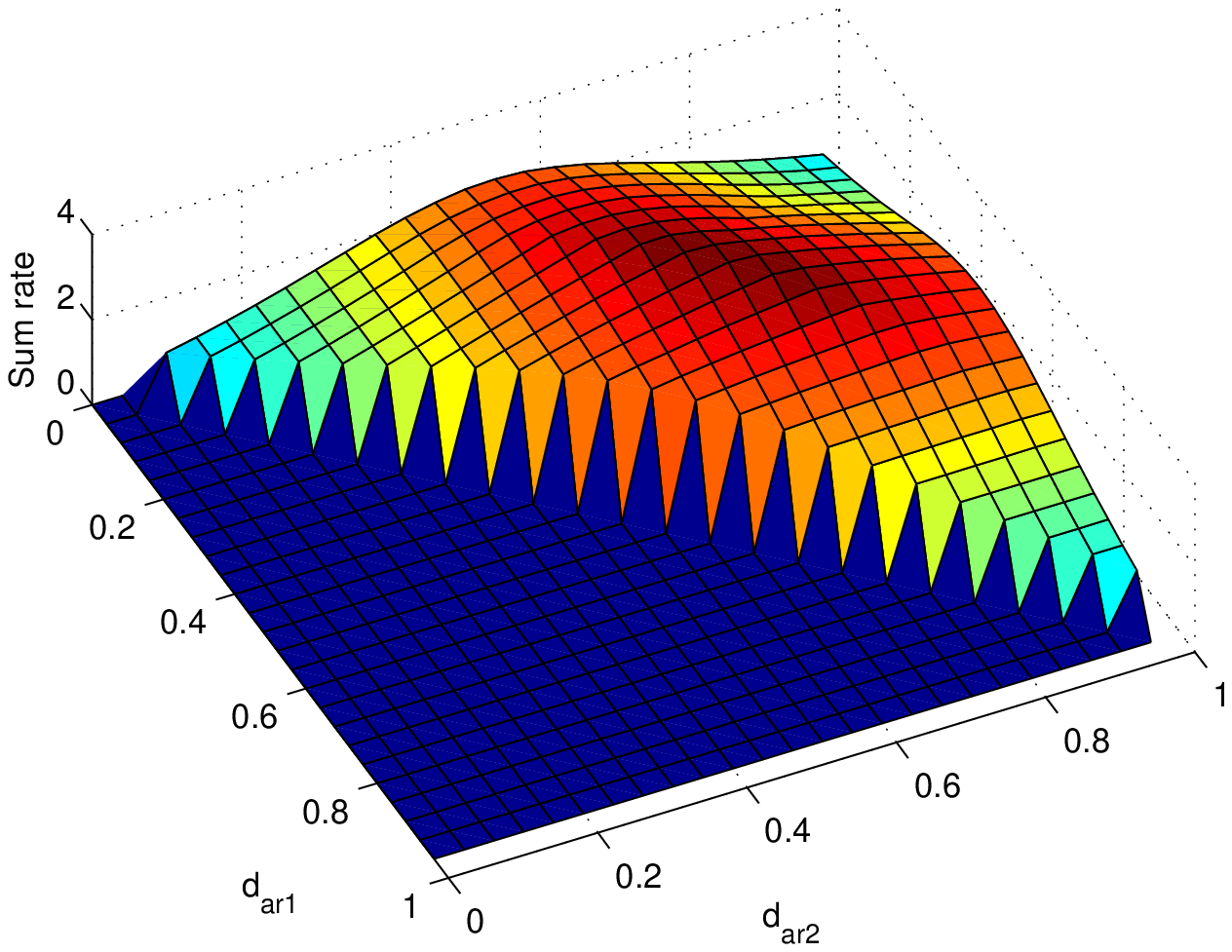, width=8cm}
      \caption{\baselineskip=10pt achievable sum rates of the $(2,4)$ DF MHMR protocol with $P=0$ dB.}
      \label{fig:two_line_df_mhmr}
    \end{center}
  \end{minipage}
  \hfill
\end{figure*}

From these plots we may conclude that the DF protocols (\Fig \ref{fig:two_line_df_mabc}, \ref{fig:two_line_df_tdbc}, \ref{fig:two_line_df_mhmr})  generally outperform  the AF protocols (\Fig \ref{fig:two_line_af_mabc}, \ref{fig:two_line_af_tdbc}, \ref{fig:two_line_af_mhmr}). In particular, the DF MHMR protocol (\Fig \ref{fig:two_line_df_mhmr}) outperforms all the other protocols. This result is consistent with  our previous low and high SNR analysis which also indicate that the DF MHMR protocol outperforms the other protocols; the plots here are for more realistic SNR values and range over a large set of channel parameters (i.e. all relay positions on the line). In the MABC and TDBC protocols (\Fig \ref{fig:two_line_af_mabc}, \ref{fig:two_line_df_mabc}, \ref{fig:two_line_af_tdbc}, \ref{fig:two_line_df_tdbc}), we achieve the largest sum data rate when two relays are located at the middle of $\na$ and $\nb$. This can be intuitively explained by noting that,  since relays have to support the bi-directional information flow, by symmetry,  the center of the two terminal nodes should be the optimal position.  One interesting observation is that we achieve close to the  largest sum rate when one relay is located in the middle of $\na$ and $\nb$ and the other relay is located at any arbitrary position in the DF protocols (\Fig \ref{fig:two_line_df_mabc}, \ref{fig:two_line_df_tdbc}). This is because we do not restrict that each relay has to decode both messages;  as such, much of performance depends on the relay in the middle, while the other one supports this relay.
In the MHMR protocols (\Fig \ref{fig:two_line_af_mhmr}, \ref{fig:two_line_df_mhmr}) the peak does not occur for relays positioned at the middle of $\na$ and $\nb$. For example, the largest sum rate is achieved when $(d_{\na\nr_1} , d_{\na\nr_2}) = (0.2 ,0.6)$ or $(0.4, 0.8)$. This implies that there exist the non-trivial optimum relay positions for the multi hop relays both in the DF and AF schemes.

\section{Conclusion}

\label{sec:conclusion} In this paper, we proposed protocols for the
half-duplex bi-directional channel with multiple relays: the $(m,2)$
MABC protocol, the $(m,3)$ TDBC protocol and the general $(m,t)$
protocol for $m$ relays and $3<t\leq m+2$ phases. We derived
achievable rate regions as well as outer bounds for 3 half-duplex bi-directional multiple
relay protocols with decode and forward relays. We compared these
regions to those achieved by the same protocols with amplify and
forward relays in the Gaussian noise channel.

Numerical evaluations suggest that the $(m,m+2)$ DF MHMR protocol
achieves the largest rate region under simulated channel conditions. { An analytical comparison of the protocols in the very high and low SNR regimes indicates that the $(m,m+2)$ DF MHMR protocol outperforms -- from a sum-rate perspective --  the other protocols.   As
expected, for a low number of hops or at high SNR AF relaying
protocols perform well, but it rapidly degrades when the number of hops
is increased or the SNR is decreased.

\appendices


\section{Proof of Theorem \ref{theorem:TDBC_mrop}}

\label{appendix:TDBC_mrop}

{\em Random code generation: } For simplicity of exposition, we take $|{\cal Q}| = 1$ and therefore
consider distributions $p^\pa(x_\na)$, $p^\pb(x_\nb)$, $p^\pc(x_{\cal{A}\cap \cal{B}})$, $p^\pc(x_{\cal{A}\setminus \cal{B}})$ and $p^\pc(x_{\cal{B}\setminus \cal{A}})$. First we generate random
$(n\cdot \Delta_{1,n})$-length sequences $\mathbf{x}^{(1)}_\na(w_\na)$ and $(n\cdot\Delta_{2,n})$-length sequences
$\mathbf{x}^\pb_\nb(w_\nb)$ i.i.d. according to  $p^\pa(x_\na)$ and $p^\pb(x_\nb)$ respectively. We also generate random $(n\cdot\Delta_{3,n})$-length sequences
${\bf x}^\pc_{\cal{A}\cap \cal{B}}(w_\nr)$ with $w_\nr \in \mathbb{Z}_L$ ($L = \max(\lfloor 2^{nR_{\na}} \rfloor, \lfloor 2^{nR_{\nb}} \rfloor )$), ${\bf x}^\pc_{\cal{A}\setminus \cal{B}}(w_\na)$ and ${\bf x}^\pc_{\cal{B}\setminus \cal{A}}(w_\nb)$, according to $p^\pc(x_{\cal{A}\cap \cal{B}})$, $p^\pc(x_{\cal{A}\setminus \cal{B}})$ and $p^\pc(x_{\cal{B}\setminus \cal{A}})$ respectively.

{\em Encoding: } During phase 1, the encoder of node $\na$ sends the codeword ${\bf x}^\pa_\na(w_\na)$. Node $\nb$ similarly sends the codeword ${\bf x}^\pa_\nb(w_\nb)$  in phase 2. Relays in ${\cal A}\cap {\cal B}$ estimate $\hat{w}_\na$ and $\hat{w}_\nb$ after phase 1 and 2 using jointly typical decoding, then construct $w_{\nr} =\hat{w}_\na \oplus \hat{w}_\nb$ in ${\mathbb Z}_L$ and send
${\bf x}^\pc_{{\cal A}\cap {\cal B}}(w_\nr)$ during phase 3. Likewise relays in ${\cal A}\setminus {\cal B}$ (resp. ${\cal B}\setminus {\cal A}$) estimate $\hat{w}_\na$ (resp. $\hat{w}_\nb$) after phase 1 (resp. phase 2) and send ${\bf x}^\pc_{{\cal A}\setminus {\cal B}}({\hat w}_\na)$ (resp. ${\bf x}^\pc_{{\cal B}\setminus {\cal A}}({\hat w}_\nb)$).

{\em Decoding: }  $\na$ estimates $\tilde{w}_\nb$ after phase 3
if $({\bf x}_\nb^\pb({\tilde w}_\nb),{\bf y}^\pb_\na)\in A^\pb(X_\nb Y_\na)$ and \\
$({\bf x}_{{\cal A}\cap{\cal B}}^\pc(w_\na \oplus {\tilde w}_\nb),{\bf x}_{{\cal B}\setminus{\cal A}}^\pc({\tilde w}_\nb),{\bf y}^\pc_\na)\in A^\pc(X_{\cal B} Y_\na)$.  Otherwise an error is declared. In ${\cal B}$, there are two different
messages $w_\nr (= w_\na \oplus w_\nb) \in {\cal A}\cap {\cal B}$
and $w_\nb \in {\cal B} \setminus {\cal A}$. However, since $\na$
knows $w_\na$ as side information $w_\nr$ is equivalent to $w_\nb$
for $\na$. Similarly, $\nb$ decodes ${\tilde
w}_\na$ after phase 3.

{\em Error analysis: }
\begin{align}
  P[E_{\nb,\na}] & \leq P[E_{\{\nb\},{\cal B}}^\pb \cup E_{\{\nb\}\cup{\cal B},\{\na\}}^\pc]\\
  & \leq P[E_{\{\nb\},{\cal B}}^\pb] + P[E_{\{\nb\}\cup{\cal B},\{\na\}}^\pc | {\bar E}_{\{\nb\},{\cal B}}^\pb ]
\end{align}
\begin{align}
P[E_{\{\nb\},{\cal B}}^\pb]
&\leq \sum_{\nr \in {\cal B}}  P[E_{\nb,\nr}^\pb]\\
&\leq \sum_{\nr \in {\cal B}}  P[\bar{D}^\pb({\bf x}_\nb(w_\nb),{\bf y}_\nr)] + 2^{nR_\nb} 2^{-n\cdot\Delta_{2,n}(I(X_\nb^\pb;Y_\nr^\pb)-3\epsilon)} \label{eq:TDBC_mrop1}\\
 P[E_{\{\nb\}\cup{\cal B},\{\na\}}^\pc | {\bar E}_{\{\nb\},{\cal B}}^\pb ]
 &\leq P[\bar{D}^\pb({\bf x}_\nb(w_\nb),{\bf y}_\na)]+ P[\bar{D}^\pc({\bf x}_{{\cal A}\cap{\cal B}}(w_\na \oplus w_\nb),{\bf x}_{{\cal B}\setminus{\cal A}}(w_\nb),{\bf y}_\na)] + \nonumber\\
 &~~~~P[\cup_{{\tilde w}_\nb \neq w_\nb} D^\pb({\bf x}_\nb({\tilde w}_\nb),{\bf y}_\na)\cap D^\pc({\bf x}_{{\cal A}\cap{\cal B}}(w_\na \oplus {\tilde w}_\nb),{\bf x}_{{\cal B}\setminus{\cal A}}({\tilde w}_\nb),{\bf y}_\na)]\\
 &\leq 2\epsilon + 2^{nR_\nb} 2^{-n(\Delta_{2,n}I(X_\nb^\pb;Y_\na^\pb) + \Delta_{3,n}I(X_{\cal B}^\pc;Y_\na^\pc) -6\epsilon)}\label{eq:TDBC_mrop2}
\end{align}

Since $\epsilon > 0$ is arbitrary, the conditions of Theorem
\ref{theorem:TDBC_mrop} and the AEP property will guarantee that the
right hand sides of \eqref{eq:TDBC_mrop1} and \eqref{eq:TDBC_mrop2}
vanish as $n \rightarrow \infty$. Similarly, $P[E_{\na,\nb}]
\rightarrow 0$ as $n \rightarrow \infty$.
By  Fenchel-Bunt's extension of Carath\'{e}odory
theorem in \cite{Hiriart:2001}, it is sufficient to restrict $|{\cal
Q}| \leq 2$.


\section{Proof of Theorem \ref{theorem:DFMH}}

\label{appendix:DFMH}

{\em Random code generation: } $\na ( = \nr_0)$ and $\nb (=\nr_{m+1})$ divide $w_\na$ and $w_\nb$ into $B$ blocks respectively. Then $\na$ has a message set $\{w_{\na,(0)},w_{\na,(1)},\cdots,w_{\na,(B-1)}\}$ and $\nb$ has a message set $\{w_{\nb,(0)},w_{\nb,(1)},\cdots,w_{\nb,(B-1)}\}$.
We generate random $(n\cdot \Delta_{(m+2-i),n})$-length sequences ${\bf x}^{(m+2-i)}_{\nr_i}(w_\nr)$ with $w_\nr \in \mathbb{Z}_L$ ($L = \max(\lfloor 2^{nR_{\na}} \rfloor, \lfloor 2^{nR_{\nb}} \rfloor )$), according to $p^{(m+2-i)}(x_{\nr_i})$ for $i \in [0,m+1]$.

{\em Encoding: }  We divide the total time period into $B+m$ time
slots. Each time slot consists of $m+2$ phases. Node $\na$ transmits
${\bf x}^{(m+2)}_\na (w_{\na,(j-1)})$ during slot $j$ and phase
$m+2$, where $j\in [1,B]$ and node $\nb$ transmits ${\bf
x}^\pa_\nb(w_{\nb,(j-m-1)})$ during slot $j$ and phase $1$, where
$j\in [m+1,B+m]$. Intermediate node $\nr_i$ ($i\in[1,m]$) transmits
${\bf x}^{(m+2-i)}_{\nr_i}(w_{\nr_i,(j)})$ during slot $j$ and phase
$m+2-i$, where $j\in[1,B+m]$ and $i\in [1,m]$. $w_{\nr_i,(j)}$ is
defined as follows:
\begin{align}
w_{\nr_i,(j)} \eqdef \left\{
                     \begin{array}{ll}
                      \tilde{w}_{\na,(j-i-1)}\oplus \tilde{w}_{\nb,(j-m-1)}, &( 1\leq j-i \leq B ~,~ m+1\leq j\leq B+m) \\
                      \tilde{w}_{\na,(j-i-1)}, & (1\leq j-i \leq B ~,~ 1 \leq j\leq m) \\
                      \tilde{w}_{\nb,(j-m-1)}, & (j-i > B  ~,~m+1\leq j\leq B+m) \\
                      \emptyset, & (1> j-i~,~ 1\leq j \leq m)
                     \end{array}
                   \right.
\end{align}

{\em Decoding: } After slot $j$ and phase $(m+2-i)$, $\nr_{i+1}$ decodes $\tilde{w}_{\na,(j-i-1)}$ using $i$ independent message lists  $L_{\nr_0,\nr_{i+1}}^{(m+2)},\cdots,L_{\nr_i,\nr_{i+1}}^{(m+2-i)}$, when $j\in [i,B+i-1]$. $L_{i,j}^{(\ell)}$ is defined as the set of the messages $w_{i,j}$ whose codewords
${\bf x}_{i}^{(\ell)}(w_{i,j})$ are jointly typical with ${\bf
y}_j^{(\ell)}$ in phase $\ell$, i.e., $L_{i,j}^{(\ell)} \eqdef \{w_{i,j}| ({\bf x}_{i}^{(\ell)}(w_{i,j}),{\bf
y}_j^{(\ell)})\in A^{(\ell)}(X_i Y_j)\}$. If there is a unique ${\tilde w}_{\na,(j-i-1)}$ in $\cap_{k=0}^i L_{\nr_k,\nr_{i+1}}^{(m+2-k)}$, $\nr_{i+1}$ declares it as the decoded message. Otherwise an error is declared. Similarly, $\nr_{i-1}$ decodes $\tilde{w}_{\nb,(j-m-1)}$ using $m+2-i$ independent message lists  $L_{\nr_i,\nr_{i-1}}^{(m+2-i)},\cdots,L_{\nr_{m+1},\nr_{i-1}}^{(1)}$, when $j\in [m+1,B+m]$. If there is a unique ${\tilde w}_{\nb,(j-m-1)}$ in $\cap_{k=i}^{m+1} L_{\nr_k,\nr_{i-1}}^{(m+2-k)}$, $\nr_{i-1}$ declares it as the decoded message. However, after phase 1 (resp. phase $m+2$), $\nr_m$ (resp. $\nr_1$) only decodes $\tilde{w}_{\nb,j}$ (resp. $\tilde{w}_{\na,j}$).

{\em Error analysis: }
\begin{align}
  P[E_{\na,\nb}] & \leq P[\cup_{i=1}^{m+1} E_{\{\nr_0,\cdots \nr_{i-1}\},\{\nr_i\}}^{(m+3-i)}]\\
&\leq \sum_{i=1}^{m+1} P[E_{\{\nr_0,\cdots \nr_{i-1}\},\{\nr_i\}}^{(m+3-i)} | \cap_{j=1}^{i-1} \bar{E}_{\{\nr_0,\cdots \nr_{j-1}\},\{\nr_j\}}^{(m+3-j)}]
\end{align}
for $i\in [1,m+1]$, we have
\begin{align}
P[E_{\{\nr_0,\cdots \nr_{i-1}\},\{\nr_i\}}^{(m+3-i)} | \cap_{j=1}^{i-1} & \bar{E}_{\{\nr_0,\cdots \nr_{j-1}\},\{\nr_j\}}^{(m+3-j)}]\nonumber\\
\leq&  \sum_{j=0}^{i-1} P[\bar{D}^{(m+2-j)}({\bf x}_{\nr_j}(w_{\nr_j}),{\bf y}_{\nr_i})] + \nonumber \\
&\sum_{{\tilde w}_\na \neq w_\na} \prod_{j=0}^{i-1} P[D^{(m+2-i)}({\bf x}_{\nr_j}({\tilde w}_\na \oplus w_\nb),{\bf y}_{\nr_i})]\\
\leq& i\cdot \epsilon + 2^{nR_\na} 2^{-n(\sum_{j=0}^{i-1} \Delta_{(m+2-j),n}I(X_{\nr_j}^{(m+2-j)};Y_{\nr_i}^{(m+2-j)}) -\epsilon')}\label{eq:DFMH1}
\end{align}

Since $\epsilon > 0$ is arbitrary, the conditions of Theorem
\ref{theorem:DFMH} and the AEP property will guarantee that the
right hand side of \eqref{eq:DFMH1}
vanishes as $n \rightarrow \infty$. Similarly, $P[E_{\nb,\na}]
\rightarrow 0$ as $n \rightarrow \infty$.


\section{Proof of Theorem \ref{theorem:TDBC:out}}

\label{appendix:TDBC:out}

We use Lemma~\ref{lemma:out} to prove the
Theorem~\ref{theorem:TDBC:out}. For every $S_R\subseteq {\cal R}$,
we have 4 kinds of cut-sets, $S_1 =\{\na\} \cup S_R$, $S_2 =\{\nb\}
\cup S_R$, $S_3 = \{\na,\nb\}\cup S_R$ and $S_4 = S_R$, as well as
two rates $R_\na$ and $R_\nb$. Also,  in the TDBC protocol,
\begin{align}
 Y_\na^\pa &= X_\nb^\pa = X_{\cal R}^\pa = \varnothing \label{TDBC:out:1} \\
 X_\na^\pb &= Y_\nb^\pb = X_{\cal R}^\pb = \varnothing \label{TDBC:out:2} \\
 X_\na^\pc &= X_\nb^\pc = Y_{\cal R}^\pc = \varnothing \label{TDBC:out:3}
\end{align}

The corresponding outer bounds for a given subset $S_R$ are :
\begin{align}
S_1 &: R_\na \leq \Delta_1 I(X_\na^\pa; Y_{\bar{S}_R}^\pa
, Y_\nb^\pa)+\Delta_3 I(X_{S_R}^\pc;Y_\nb^\pc|X_{\bar{S}_R}^\pc)+ \epsilon, \label{TDBC:out:4}\\
S_2 &: R_\nb \leq \Delta_2 I(X_\nb^\pb; Y_{\bar{S}_R}^\pb
,Y_\na^\pb)+\Delta_3
I(X_{S_R}^\pc;Y_\na^\pc|X_{\bar{S}_R}^\pc)+\epsilon,
\label{TDBC:out:5}
\end{align}

The cut-sets $S_3$ and $S_4$ yield no constraints.
Since $\epsilon > 0$ is arbitrary, \eqref{TDBC:out:4} and
\eqref{TDBC:out:5} yields the Theorem~\ref{theorem:TDBC:out}.

\bibliographystyle{IEEEtranS}
\bibliography{sang}
\end{document}